 \documentclass[preprint,showpacs,preprintnumbers,amsmath,amssymb, superscriptaddress,aps,prd,longbibliography,nofootinbib]{revtex4-1}
\usepackage{graphicx} 
\usepackage{xcolor}
\usepackage{amsmath}
\usepackage{hyperref}
\usepackage{subfigure}
\usepackage{amsfonts}
\usepackage[top = 2cm, bottom = 2cm, right = 2cm, left =2cm]{geometry}
\usepackage{amssymb}
\graphicspath{ {figs/} }
\usepackage[latin1]{inputenc}
\usepackage[english]{babel}
\usepackage{epsfig}
\usepackage{wasysym}
\usepackage{color,xcolor}

\begin{document}

\title{Spherically symmetric and static black bounces with multiple horizons, throats, and anti-throats in four dimensions}

 \author{Manuel E. Rodrigues} 
 \email{esialg@gmail.com}
\affiliation{Faculdade de F\'{i}sica, Programa de P\'{o}s-Gradua\c{c}\~{a}o em F\'{i}sica, Universidade Federal do Par\'{a}, 66075-110, Bel\'{e}m, Par\'{a}, Brazill}
\affiliation{Faculdade de Ci\^{e}ncias Exatas e Tecnologia, Universidade Federal do Par\'{a}, Campus Universit\'{a}rio de Abaetetuba, 68440-000, Abaetetuba, Par\'{a}, Brazil}

\author{Marcos V. de S. Silva \footnote{Author to whom any correspondence should be addressed.}}
\email{marco2s303@gmail.com}
\affiliation{Departamento de F\'isica, Universidade Federal do Cear\'a, Caixa Postal 6030, Campus do Pici, 60455-760 Fortaleza, Cear\'a, Brazil}

\date{\today}

\begin{abstract}
Black bounce spacetimes usually arise from the Simpson-Visser regularization method. This type of metric presents a wormhole throat inside an event horizon. In this paper, we presented new classes of black bounce spacetime solutions, which have multiple horizons, throats, and anti-throats. These solutions are variants of black holes and wormholes, based on modifications of the Schwarzschild and Simpson-Visser metrics. The metric function allows for multiple horizons and throats, and the asymptotic behavior recovers the Schwarzschild solution. The article considers a scalar field coupled to nonlinear electrodynamics, generating solutions with a partially phantom scalar field and a magnetic monopole. The energy conditions can be satisfied or violated depending on the region of spacetime, analyzed through an anisotropic fluid. The regularity of the spacetime is ensured by the analysis of the Kretschmann scalar. 
\end{abstract}

 \maketitle

\section{Introduction}\label{S:Introduction}
Black holes are compact objects that originally arise as solutions to Einstein equations in general relativity \cite{Wald:1984rg}. These objects are ideal candidates for studying gravity in the strong field regime, allowing us to analyze nonlinear effects of the theory or to evaluate deviations from general relativity \cite{Capozziello:2011et}. Another important aspect of black holes is their causal structure. These objects possess a region of no return, called the event horizon, which causally separates events occurring inside and outside the black hole \cite{Wald:1984rg}. It is important to note that the main characteristic of these objects is the presence of an event horizon. Depending on the type of black hole, the number of horizons can be indefinite \cite{Ansoldi:2012qv}.

A major issue in general relativity arises when studying the internal structure of black holes. Standard black holes contain a singularity hidden within the event horizon. The singularity can be understood as a point, or a set of points, where geodesics are interrupted \cite{Bronnikov:2012wsj}. In alternative theories of gravity, the problem of singularities can be circumvented \cite{Olmo:2015axa}. In the context of general relativity, the issue of singularities can be avoided by introducing exotic fields. For example, phantom scalar fields can generate solutions without singularities \cite{Bronnikov:2005gm}. Black holes that do not exhibit singularities are known as regular black holes \cite{Ansoldi:2008jw}. The first regular model was proposed by Bardeen in 1968, and it was derived from Einstein equations by Beato and Garcia \cite{Ayon-Beato:2000mjt}.

There are different types of regular black holes. Regular black holes, such as the Bardeen model, are said to have a regularized interior, where the center can be either de Sitter-like or Minkowski-like \cite{Ansoldi:2008jw,Simpson:2019mud}. Solutions with a de Sitter center exhibit a $g_{00}\approx 1-r^2$ behavior and have a constant energy density in the center, while solutions with a $g_{00}\approx1-r^n$ behavior, with $n\geq3$, are Minkowski-like and feature an energy density that tends to zero at the center \cite{Bronnikov:2024izh}. In the literature, there are various examples of these types of solutions, with different properties being studied. Recently, Bronnikov proposed a regularization method capable of generating this type of solution by regularizing the center of known singular solutions \cite{Bronnikov:2024izh,Bolokhov:2024sdy}.

Simpson and Visser proposed another type of regular spacetime in 2018 \cite{Simpson:2018tsi}. This regular black hole, known as black bounce (BB), differs from usual regular black holes because they feature a wormhole structure hidden inside the event horizon. Although they can be constructed in other ways \cite{Lobo:2020ffi}, BB spacetimes usually arise through the Simpson-Visser regularization \cite{Franzin:2021vnj,Furtado:2022tnb,Lima:2022pvc,Bronnikov:2023aya,Lima:2023arg,Lima:2023jtl,Crispim:2024yjz}, which removes the singularities but introduces a wormhole structure within the horizon of these solutions. These metrics are capable of closely mimicking the shadows of singular black holes \cite{Lima:2021las,Guerrero:2021ues,Vagnozzi:2022moj}.

A problem that arises with regularized metrics is identifying the matter content capable of generating these objects. For example, the Schwarzschild solution is a vacuum solution, whereas its regularized version, described by the Simpson-Visser spacetime, does not satisfy the Einstein equations for vacuum \cite{Simpson:2018tsi}. In various works, Bronnikov demonstrated how the source of these solutions can be determined by analyzing the symmetries of the Einstein tensor \cite{Bronnikov:2021uta,Bronnikov:2022bud,Bronnikov:2023aya,Bronnikov:2024izh,Bolokhov:2024sdy}. He showed that solutions with regular centers, whether of the de Sitter or Minkowski type, can be constructed using only nonlinear electrodynamics (NED), while BB solutions require both NED and a phantom scalar field.

In addition to regularizing solutions, NED alters important properties of black holes. Various studies have shown that NED modifies black hole thermodynamics by altering the first law of thermodynamics \cite{Ma:2014qma,Zhang:2016ilt,Maluf:2018lyu,Rodrigues:2022qdp,deSSilva:2024fmp}. Photons do not follow null geodesics due to high interaction terms in the presence of NED \cite{Bronnikov:2000vy,Novello:1999pg,Toshmatov:2021fgm,Bronnikov:2022ofk,Silva:2024fpn}, so the shadow of a black hole is modified if the source of the solution is a nonlinear charge \cite{Stuchlik:2019uvf,Rayimbaev:2022znx,dePaula:2023ozi,daSilva:2023jxa}. Black holes with multiple horizons, whether singular or regular, can also originate from NED \cite{Nojiri:2017kex,Gao:2017vqv,Rodrigues:2019xrc,Rodrigues:2020pem,Nashed:2021ctg}. Thus, NED plays a crucial role in the study of these solutions.

The structure of this article is organized as follows: In Sec. \ref{S:Reg_Sol}, we present the multihorizon solution in $3+1$ dimensions and discuss the properties of the spacetime, such as regularity, horizons, and throats/anti-throats. The field sources are studied in Sec. \ref{S:Field_Source}. In Sec. \ref{S:EC}, we examine which of the energy conditions are violated. Our conclusions and perspectives are presented in Sec. \ref{S:conclusions}.

\section{The black bounce spacetime}\label{S:Reg_Sol}
The BB spacetimes are characterized by the presence of a type of wormhole located within an event horizon. This type of object can be described by a line element of the form \cite{Bronnikov:2021uta}:
\begin{equation}
    ds^2=A(x)dt^2-\frac{1}{A(x)}dx^2-r^2(x)\left(d\theta^2+\sin^2\theta d\varphi^2\right).\label{line}
\end{equation}
Horizons are null hypersurfaces that causally separate distinct regions of spacetime. For spherically symmetric, static, and asymptotically flat spacetimes, the horizons (not necessarily the event horizon) can be obtained through the condition $g_{00} = 0$ \cite{Morris:1988cz,Vishveshwara1968}, which in the cases we are considering coincide with the Killing horizon of our spacetime, and thus we obtain the horizons of this solution by solving $A(x) = 0$. The function $r(x)$ must have nonzero minimum values to ensure the presence of wormhole throats.

Our aim is to obtain BB solutions with a richer causal structure than that of usual BBs. Thus, we want our model to possess an undetermined number of horizons, throats, and anti-throats. We also require that our model be a generalization of the Simpson-Visser proposal, so that, in some limit, our solution recovers the Simpson-Visser solution and, consequently, recovers Schwarzschild, in some limit, and is asymptotically flat. Thus, we will consider the model described by:
\begin{equation}
    A(x)=1-\frac{2 M \cos \left(\frac{ a_0}{\sqrt{a^2+x^2}}\right)}{\sqrt{a^2+x^2}}, \quad \mbox{and} \quad r^2(x)=x^2 \cos ^2\left(\frac{  a_0}{\sqrt{a^2+x^2}}\right)+a^2.\label{metric_func}
\end{equation}
The parameters $a$ and $a_0$ will determine the number of horizons and throats/anti-throats that this spacetime can exhibit. If the parameter $a_0=0$, we recover the Simpson-Visser solution. If $a_0=a=0$, the Schwarzschild solution is obtained again.

If we expand the metric coefficients, we find
\begin{equation}
    A(x)\approx 1-\frac{2 M}{x}+O\left(\frac{1}{x^3}\right), \quad \mbox{and} \quad r\approx x +O\left(\frac{1}{x}\right).
\end{equation}
This means that asymptotically we recover the Schwarzschild solution.

\begin{figure}
    \centering
    \includegraphics[width=0.45\linewidth]{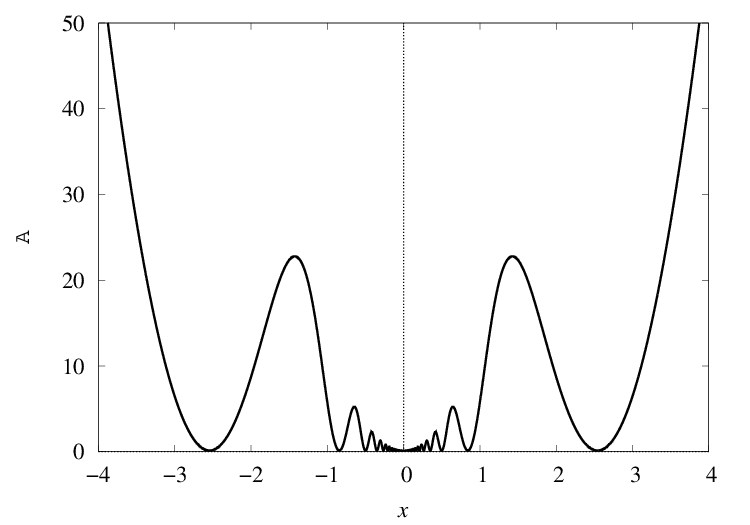}
    \includegraphics[width=0.45\linewidth]{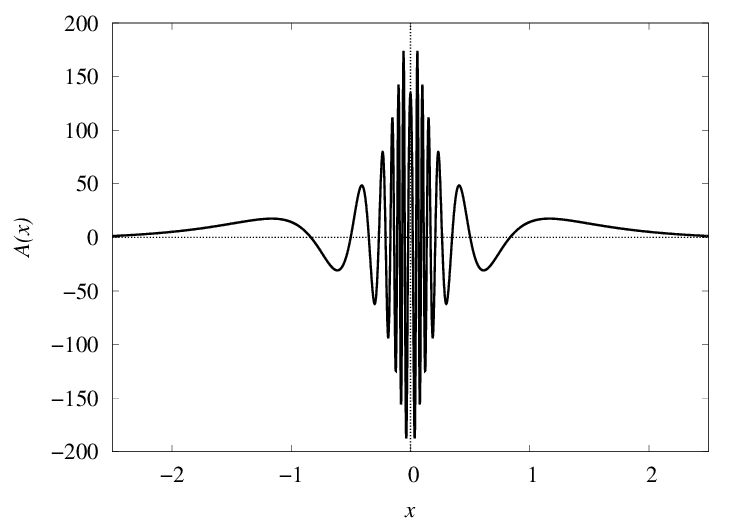}
    \caption{Graphical representation of the area of a spherical surface with radius $x$ and the function 
$A(x)$. For the graphs, we chose $a_0=4$, $a=0.1$, and $M=10$. Depending on the values chosen for these parameters, the number of horizons, throats, and anti-throats, as well as their positions, may vary.}
    \label{fig:areaf}
\end{figure}

The area of a sphere at radial coordinate $x$ is now
\begin{equation}
    \mathcal{A}=4\pi r^2(x)=4 \pi \left[x^2 \cos ^2\left(\frac{ a_0}{\sqrt{a^2+x^2}}\right)+a^2\right].
\end{equation}
In Fig. \ref{fig:areaf}, we can observe the behavior of the area and the function $A(x)$, noting that these functions oscillate several times with the radial coordinate. Each zero of the function $A(x)$ represents the presence of a different horizon, the minima in the area represent wormhole throats, and the maxima in the area represent wormhole anti-throats \cite{Rodrigues:2022mdm}.

\begin{figure}
    \centering
    \includegraphics[width=0.5\linewidth]{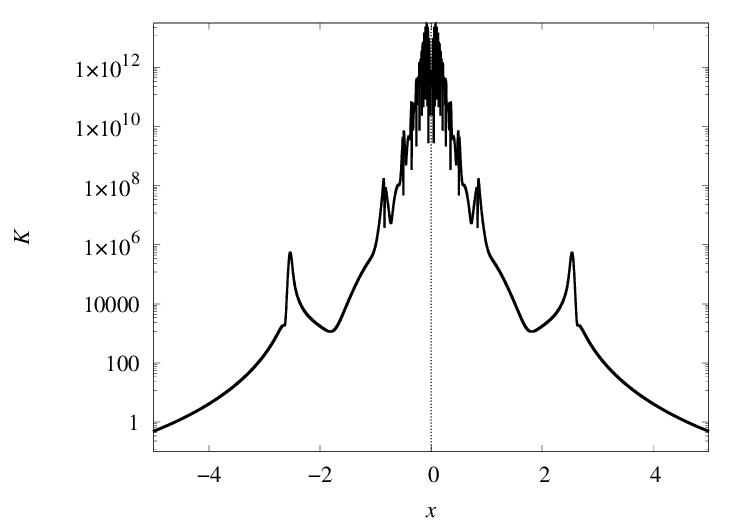}
    \caption{Graphical representation of the Kretschmann scalar. For the graph, we chose $a_0 = 4$, $a = 0.1$, and $M = 10$.}
    \label{fig:kre}
\end{figure}

In order to verify the regularity of the spacetime, we may also calculate the Kretschmann scalar, $K=R_{\alpha\beta\mu\nu}R^{\alpha\beta\mu\nu}$, which is written as~\cite{Lobo:2020ffi}
\begin{eqnarray}
&&K=\frac{
(r^2 A'')^2+2(r A' r')^2 +2r^2(A' r'+2A r'')^2
+4(1-A r'^2)^2}
{r^4}\,.
\label{Kret3}
\end{eqnarray}
For an asymptotically flat spacetime, the functions should behave as $\lim_{x\to\infty} \{A(x),r(x)/x\}=\{1,1\}$. In this limit, the Kretschmann scalar is null. The expression of the Kretschmann scalar to our solution is complicated. So that, we only analyze the asymptotic behaviors, which are given by:
\begin{eqnarray}
    K(x\rightarrow 0) &\approx& \frac{4 M^2 \left(a \cos \left(\frac{a_0}{a}\right)-a_0 \sin
   \left(\frac{a_0}{a}\right)\right)^2}{a^8}+\frac{8 \cos ^4\left(\frac{a_0}{a}\right) \left(a-2 M
   \cos \left(\frac{a_0}{a}\right)\right)^2}{a^6}+\frac{4}{a^4}+O\left(x^2\right),\nonumber\\\\
   K(x\rightarrow \infty) &\approx& \frac{48 M^2}{x^6}+\frac{32 M (a-a_0) (a+a_0)}{x^7}+O\left(\frac{1}{x^8}\right).
\end{eqnarray}
This mean that the spacetime is regular at the infinity and in the center. To verify if our spacetime is regular in all regions, we analyze the Kretschmann scalar graphically, Fig. \ref{fig:kre}. For the chosen parameters, these regions exhibit a series of horizons, throats, and anti-throats, as shown in Fig. \ref{fig:areaf}. Thus, we verify that the Kretschmann scalar does not present singularities at any point, including the horizons, throats, and anti-throats.

Following the proceeding used in \cite{Lobo:2020ffi}, we may also calculate the Hernandez-Misner-Sharp quasi-local mass, which is the energy contained within a surface of radius $x$, as
\begin{equation}
    M_{HMS}(x)=\frac{1}{2}r(x)\left\{1-A(x) r'(x)^2\right\}.
\end{equation}
Using the metric functions given by \eqref{metric_func}, we find that the Hernandez-Misner-Sharp quasi-local mass to our model is
\begin{eqnarray}
    M_{HMS}&=&\frac{1}{2} \sqrt{x^2 \cos ^2\left(\frac{a_0}{\sqrt{a^2+x^2}}\right)+a^2} \Bigg(1-\frac{x^2\cos ^2\left(\frac{a_0}{\sqrt{a^2+x^2}}\right)
   }{x^2 \cos ^2\left(\frac{a_0}{\sqrt{a^2+x^2}}\right)
  +a^2}\times\nonumber\\
    && \left(1-\frac{2 M \cos
   \left(\frac{a_0}{\sqrt{a^2+x^2}}\right)}{\sqrt{a^2+x^2}}\right)\left(\frac{a_0
   x^2 \sin \left(\frac{a_0}{\sqrt{a^2+x^2}}\right)}{\left(a^2+x^2\right)^{3/2}}+\cos
   \left(\frac{a_0}{\sqrt{a^2+x^2}}\right)\right)^2\Bigg).\label{mass}
\end{eqnarray}
In the case $a_0=0$ we find
\begin{equation}
    M_{HMS}=\frac{M x^2}{a^2+x^2}+\frac{a^2}{2 \sqrt{a^2+x^2}},
\end{equation}
which is the Hernandez-Misner-Sharp quasi-local mass to the Simpson-Visser solution. The function \eqref{mass} has the limits $\lim_{x\rightarrow 0}M_{HMS}(x)=a/2$ and $\lim_{x\rightarrow \infty}M_{HMS}(x)=M$, which are the same limits of the models proposed in \cite{Lobo:2020ffi}.

\begin{figure}
    \centering
    \includegraphics[width=0.5\linewidth]{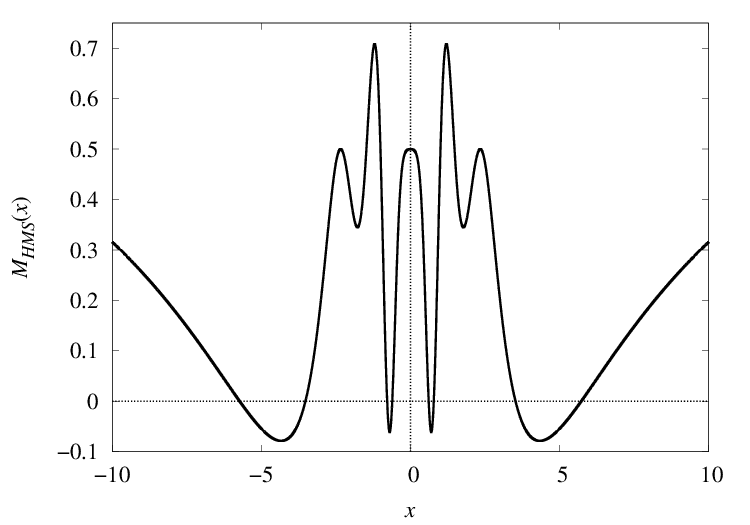}
    \caption{Hernandez-Misner-Sharp quasi-local mass for the model \eqref{metric_func}, considering the parameters $a=M=1$ and $a_0=4$.}
    \label{fig:mass}
\end{figure}

Usually, BB solutions exhibit a positive Hernandez-Misner-Sharp quasi-local mass, however this is not the case for our model. From Fig. \ref{fig:mass}, we can observe that there are cases where $M_{HMS}$ becomes negative, thus ensuring that it is not positive at all points in spacetime.

\section{Field sources}\label{S:Field_Source}
It's important to know the type of field source that can generates this spacetime. In order to obtain the field sources, we will consider the theory described by the action:
\begin{equation}
    S=\int \sqrt{\left|g\right|}d^4x\left[R - 2 h\left(\phi\right) g^{\mu\nu}\partial_\mu \phi\partial_\nu \phi
   		+2V(\phi) + L(F)\right],    			\label{Action}
\end{equation}
where $\phi$ is the scalar field, $V(\phi)$ is its potential, $L(F)$ is the NED Lagrangian, $F=F^{\mu\nu}F_{\mu\nu}$, and $F_{\mu\nu} = \partial_\mu A_\nu - \partial_\nu A_\mu$ is the electromagnetic field tensor. The function $h\left(\phi\right)$ will determine if the scalar field is phantom, $h\left(\phi\right)<0$, or standard, $h\left(\phi\right)>0$.
  
Varying the action \eqref{Action} with respect to $\phi$, $A_\mu$, and $g^{\mu\nu}$, we obtain the field equations
\begin{eqnarray}
    &&\nabla_\mu \left[L_F F^{\mu\nu}\right] =\frac{1}{\sqrt{\left|g\right|}}\partial_\mu \left[\sqrt{\left|g\right|}L_F F^{\mu\nu}\right]=0,\label{eq-F}\\
     &&2h\left(\phi\right) \nabla_\mu \nabla^\mu\phi +\frac{dh\left(\phi\right)}{d\phi}\partial^\mu\phi\partial_\mu\phi=-  \frac{dV(\phi)}{d\phi} ,     \label{eq-phi}\\  
      &&G_{\mu\nu} =R_{\mu\nu}-\frac{1}{2}R g_{\mu\nu} =T[\phi]_{\mu\nu}  + T[F]_{\mu\nu},				\label{eq-Ein}
\end{eqnarray}
where $L_F=d L/d F$, $T[\phi]_{\mu\nu}$ and $T[F]_{\mu\nu}$ are the stress-energy tensors of the scalar and electromagnetic fields, respectively,
\begin{eqnarray}                
   T[F]_{\mu\nu} = \frac{1}{2} g_{\mu\nu} L(F ) - 2L_F {F_{\nu}}^{\alpha} F_{\mu \alpha},\label{SETF}\\
    T[\phi]_{\mu\nu} =  2 h\left(\phi\right) \partial_\nu\phi\partial_\mu\phi 
   - g_{\mu\nu} \big (h\left(\phi\right) \partial^\alpha \phi \partial_\alpha \phi - V(\phi)\big).\label{SETphi}
\end{eqnarray}

Solving the Maxwell equations, considering an magnetically charged solution, we find that the magnetic field has the form
\begin{equation}
    F_{23}=q \sin \theta,
\end{equation}
and the scalar $F$ is
\begin{equation}
    F=\frac{2q^2}{ r^2(x)},
\end{equation}
where $q$ is the magnetic charge.

Considering the line element given by \eqref{line}, the field equations are given by:
\begin{eqnarray}
    -\frac{A' r'}{r}-\frac{2 A r''}{r}-\frac{A r'(x)^2}{r^2}-A h
    \phi '^2-\frac{L}{2}+\frac{1}{r^2}-V=0,\\
   \frac{1-r' \left(r A'+A r'\right)}{r^2}+A  h \phi
   '^2-\frac{L}{2}-V=0,\\
   \frac{2 q^2 L_F}{r^4}-\frac{r \left(A''+2 A h \phi '^2+L+2
   V\right)+2 A' r'+2 A r''}{2 r}=0,\\
   -2 h  \left(A' \phi '+A \phi ''\right)-\frac{4 A h  r' \phi
   '}{r}-A \phi ' h '+\frac{V'}{\phi '}=0.
\end{eqnarray}

From the field equations, we may write the electromagnetic functions as:
\begin{eqnarray}
    L(x)=-\frac{2 \left(r A' r'+A r'^2+r^2 \left(V-A h  \phi
   '^2\right)-1\right)}{r^2},\\
   L_F(x)=\frac{r^2 \left(r^2 A''+A \left(2 r \left(r''+2 r h  \phi
   '^2\right)-2 r'^2\right)+2\right)}{4 q^2}.
\end{eqnarray}
In this way, the electromagnetic expressions will only be determined once we have the expressions for the functions related to the scalar field. From the equations of motion, we can also write:
\begin{equation}
   h(x)\phi '(x)^2= -\frac{r''}{r}.\label{eqhphi}
\end{equation}
For simpler models of $r(x)$, the function $h(\phi)$ typically is given by $\pm 1$, and by integrating this equation, we obtain the scalar field. However, due to the complexity of our case, we will follow the approach used in \cite{Bronnikov:2022bud} and choose the scalar field as a monotonic function as $\phi=\arctan(x/a)$. From the equation \eqref{eqhphi}, we then obtain 
$h(\phi(x))$, which is given by:
\begin{equation}
    h(\phi(x))=-\frac{\left(a^2+x^2\right)^2 r''(x)}{a^2 r(x)}.
\end{equation}
Some of the functions can only be obtained numerically, so we will present the results graphically.

\begin{figure}
    \centering
    \includegraphics[width=0.49\linewidth]{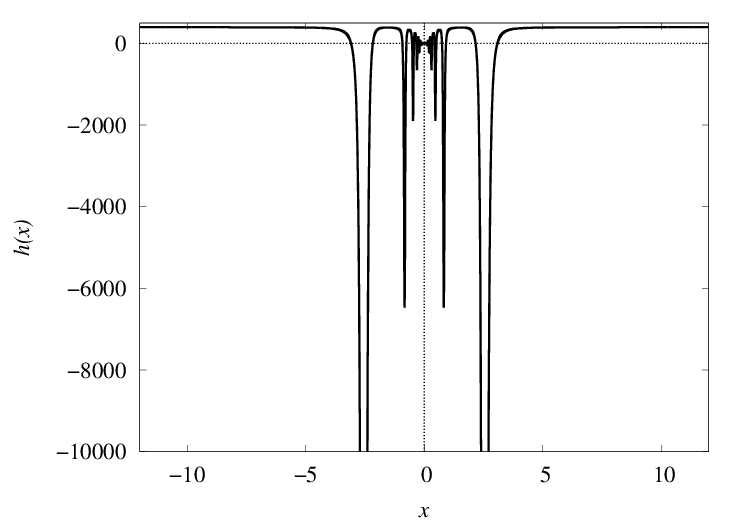}
    \includegraphics[width=0.49\linewidth]{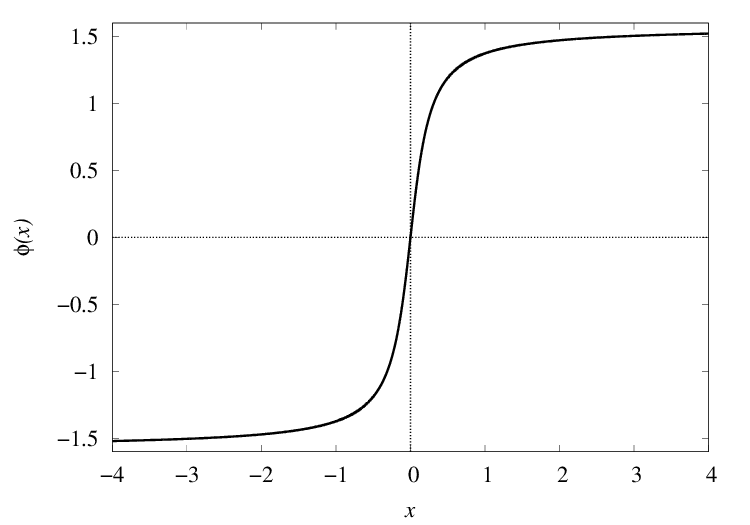}
    \caption{Behavior of the functions $h(\phi(x)$ and $\phi(x)$ in terms of the radial coordinate. We are considering $a_0=4$, $M=10$, and $a=0.2$.}
    \label{fig:hphi}
\end{figure}

\begin{figure}
    \centering
    \includegraphics[width=0.49\linewidth]{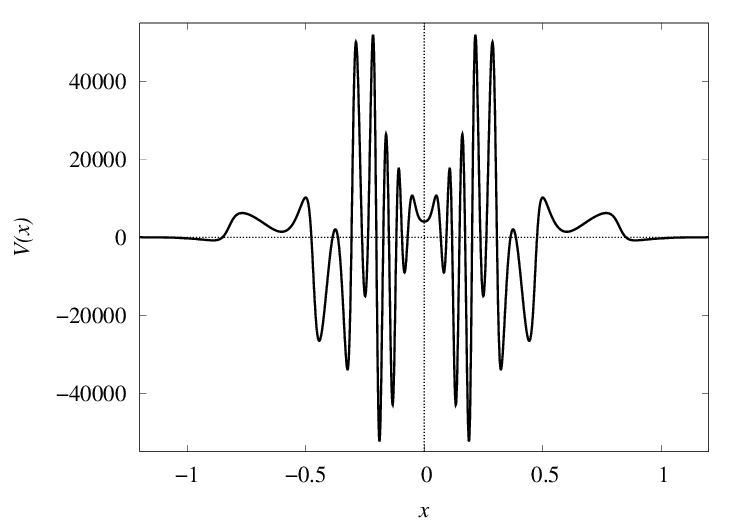}
    \includegraphics[width=0.49\linewidth]{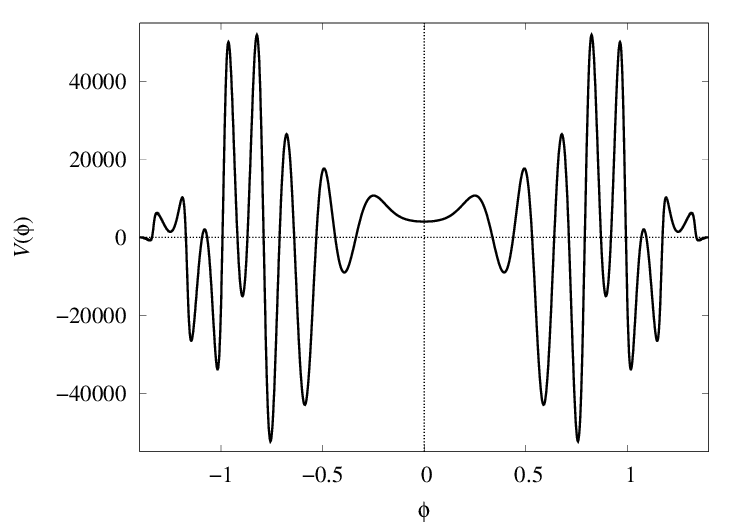}
    \caption{Potential associated with the scalar field as a function of the radial coordinate (left) and in terms of the scalar field (right). The parameters values are $a_0=4$, $M=10$, and $a=0.2$.}
    \label{fig:vx}
\end{figure}

In Fig. \ref{fig:hphi}, we analyzed the behavior of the functions. We observe that the scalar field is well-behaved, being a monotonic function, and the function that determines whether the field is canonical or phantom changes sign multiple times. In this way, at more distant points, the field behaves canonically and transits to phantom-like behavior multiple times as it approaches the center of the solution. The potential associated with the scalar field can be analyzed graphically through Fig. \ref{fig:vx}. The potential is symmetric and well-behaved, exhibiting oscillatory behavior.

\begin{figure}
    \centering
    \includegraphics[width=0.49\linewidth]{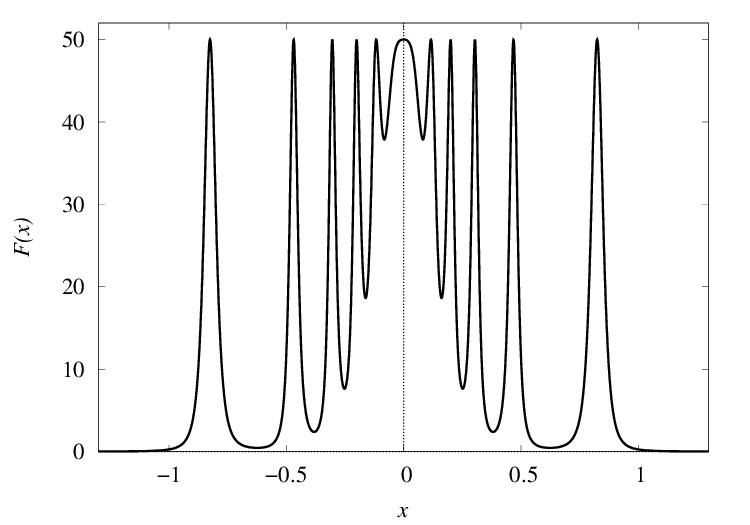}
    \includegraphics[width=0.49\linewidth]{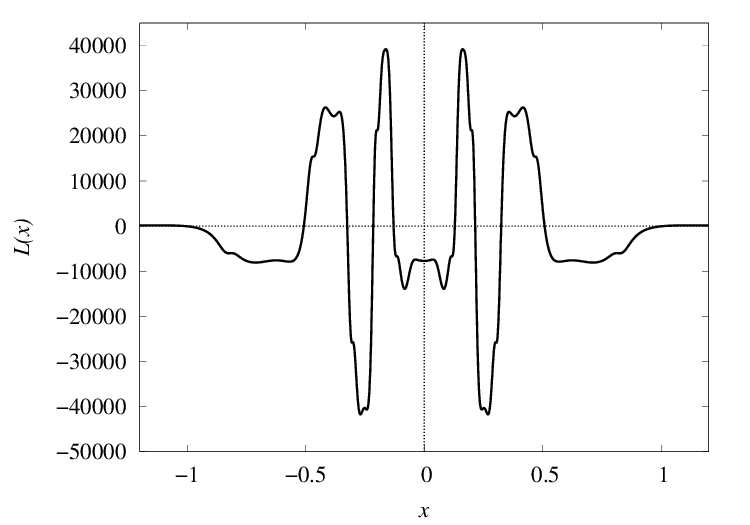}
    \includegraphics[width=0.49\linewidth]{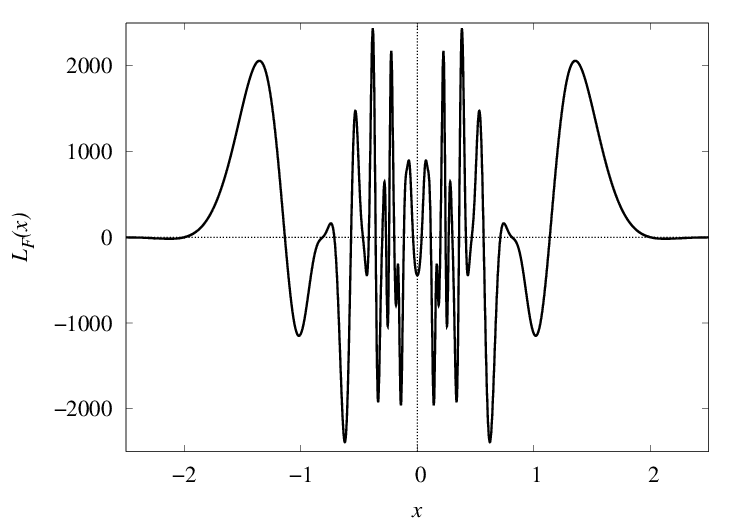}
    \caption{Behavior of the electromagnetic functions in terms of the radial coordinate with $a_0=4$, $M=10$, and $q=a=0.2$.}
    \label{fig:LFx_Lx}
\end{figure}

As expected, Fig. \ref{fig:LFx_Lx} shows that the electromagnetic functions exhibit oscillatory behavior. Although quite complex, the electromagnetic functions obey the consistency relation:
\begin{equation}
    L_F \frac{dF}{dr}-\frac{dL}{dr}=0.
\end{equation}
The oscillations in the function $F(x)$ create difficulties in representing the Lagrangian in terms of the electromagnetic scalar, $L(F)$. This happens because each oscillation in this function generates two different curves in the Lagrangian, where the transition from one curve to another through a cusp \cite{Bronnikov:2000vy}.

Thus, the behavior of all functions related to the source fields is described.
\section{Energy conditions}\label{S:EC}
In this section, we will analyze the energy conditions for our BB model. In order to do that, we rewrite the Einstein equations as
\begin{equation}
\label{EoM}
 R_{\mu\nu}-\frac{1}{2} g_{\mu\nu} R=T_{\mu\nu},
\end{equation}
where $T_{\mu\nu}$ is the stress-energy tensor that describes an anisotropic fluid. In regions where $A(x)>0$, we can write the stress-energy tensor as
\begin{equation}
    {T^{\mu}}_{\nu}=\mbox{diag}\left[\rho,\, -p_r,\, -p_t,\, -p_t\right],
\end{equation}
where $\rho$ is the energy density, $p_r$ is the radial pressure, and $p_t$ is the tangential pressure. In regions where $A(x)<0$, the stress-energy tensor is written as
\begin{equation}
    {T^{\mu}}_{\nu}=\mbox{diag}\left[-p_r,\, \rho,\, -p_t,\, -p_t\right].
\end{equation}

From the field equations \eqref{EoM}, we can immediately obtain the expressions for the fluid quantities. We have for $A(x)>0$:
\begin{equation}
\rho =-\frac{A' r'}{r}-\frac{2 A r''}{r}-\frac{A r'^2}{r^2}+\frac{1}{r^2},\quad
p_r=\frac{A' r'}{r}+\frac{A r'^2}{r^2}-\frac{1}{r^2},\quad
p_t=\frac{A''}{2}+\frac{A' r'}{r}+\frac{A r''}{r}. 
\end{equation}
For $A(x)<0$ we have:
\begin{equation}
    \rho=-\frac{A' r'}{r}-\frac{A r'^2}{r^2}+\frac{1}{r^2},\quad p_r=\frac{A' r'}{r}+\frac{2 A r''}{r}+\frac{A r'^2}{r^2}-\frac{1}{r^2}, \quad p_t=\frac{A''}{2}+\frac{A' r'}{r}+\frac{A r''}{r}.
\end{equation}

Once we have the fluid quantities, the energy conditions are given by \cite{Visser:1995cc}:
\begin{eqnarray}
&&NEC_{1,2}=WEC_{1,2}=SEC_{1,2} 
\Longleftrightarrow \rho+p_{r,t}\geq 0,\label{Econd1} \\
&&SEC_3 \Longleftrightarrow\rho+p_r+2p_t\geq 0,\label{Econd2}\\
&&DEC_{1,2} \Longleftrightarrow \rho-|p_{r,t}|\geq 0 \Longleftrightarrow 
(\rho+p_{r,t}\geq 0) \hbox{ and } (\rho-p_{r,t}\geq 0),\label{Econd3}\\
&&DEC_3=WEC_3 \Longleftrightarrow\rho\geq 0.\label{Econd4}
\end{eqnarray}
The notation we are using refers to NEC as the null energy condition, WEC as the weak energy condition, DEC as the dominant energy condition, and SEC as the strong energy condition.

Typically, BB solutions always violate the energy conditions since the null energy condition is given by:
\begin{equation}
    NEC_{1}\Longleftrightarrow -\frac{2Ar''}{r}\geq 0, \quad \mbox{for} \quad A(x)>0,\quad \mbox{and} \quad 
    NEC_{1}\Longleftrightarrow \frac{2Ar''}{r}\geq 0, \quad \mbox{for} \quad A(x)<0,
\end{equation}
and $r''/r>0$ for most existing BB solutions \cite{Alencar:2024yvh,Lobo:2020ffi}. However, to our model it's not necessarily true. The expressions for the energy conditions in our model are quite complex, but if we expand the first inequality of the null energy condition, we obtain that:
\begin{equation}
    -\frac{2Ar''}{r}\approx \frac{2 a_0^2-2 a^2}{x^4}+O\left(\frac{1}{x^5}\right), \quad \mbox{to} \quad r\rightarrow \infty.
\end{equation}
This means that depending on the relationship between the parameters, the null energy condition can be satisfied in some regions.

\begin{figure}
    \centering
    \includegraphics[width=0.49\linewidth]{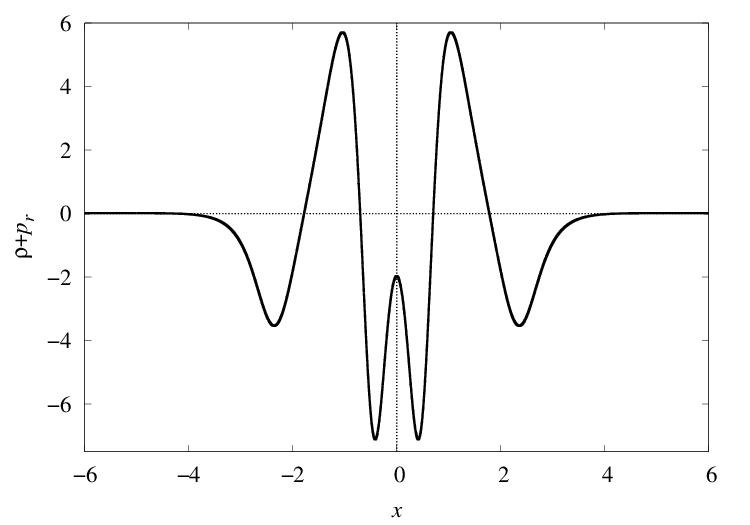}
    \includegraphics[width=0.49\linewidth]{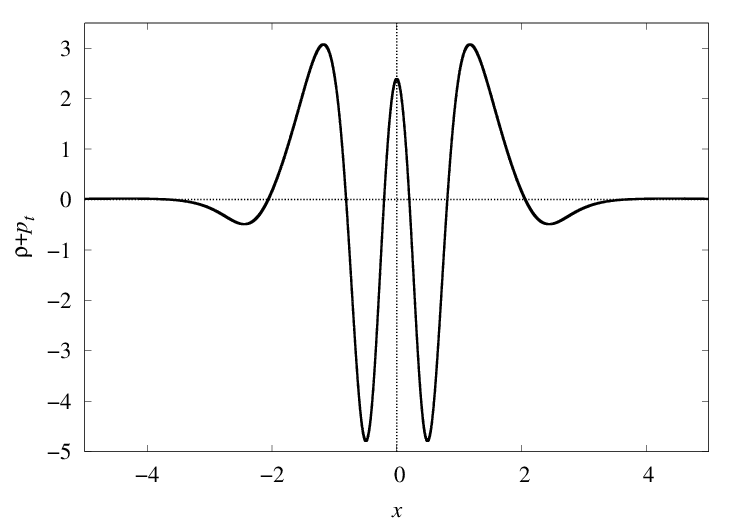}
    \includegraphics[width=0.49\linewidth]{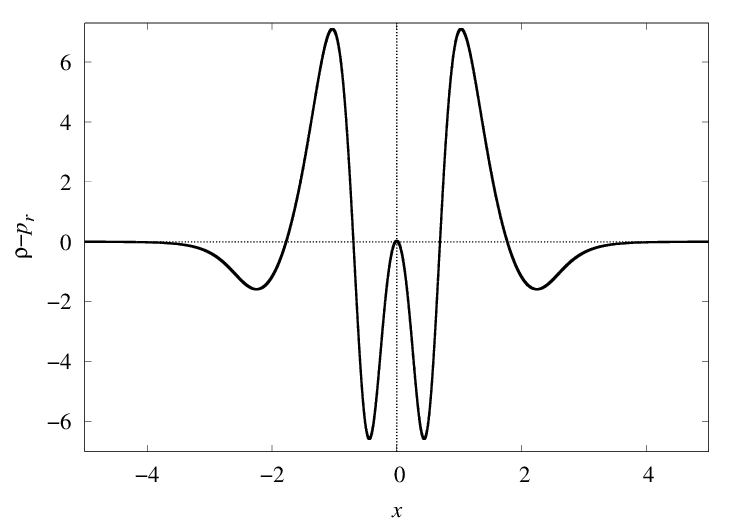}
    \includegraphics[width=0.49\linewidth]{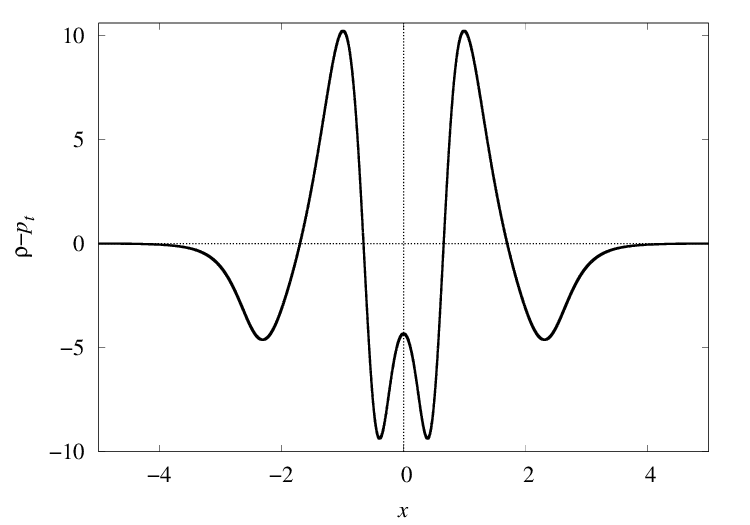}
    \includegraphics[width=0.49\linewidth]{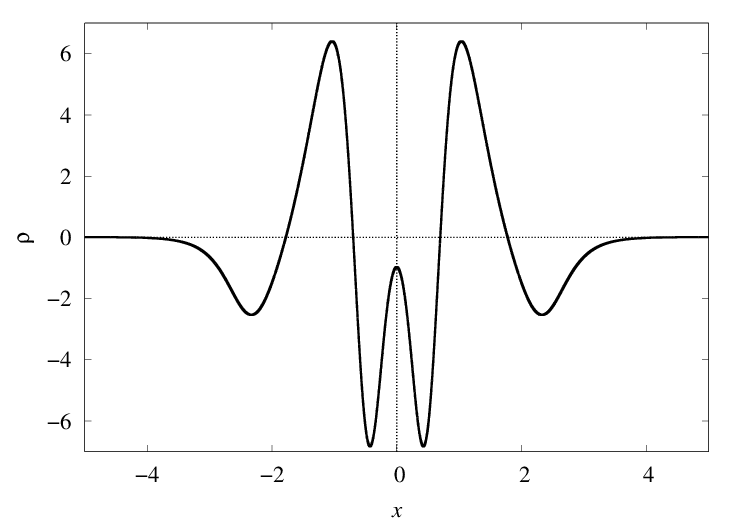}
    \includegraphics[width=0.49\linewidth]{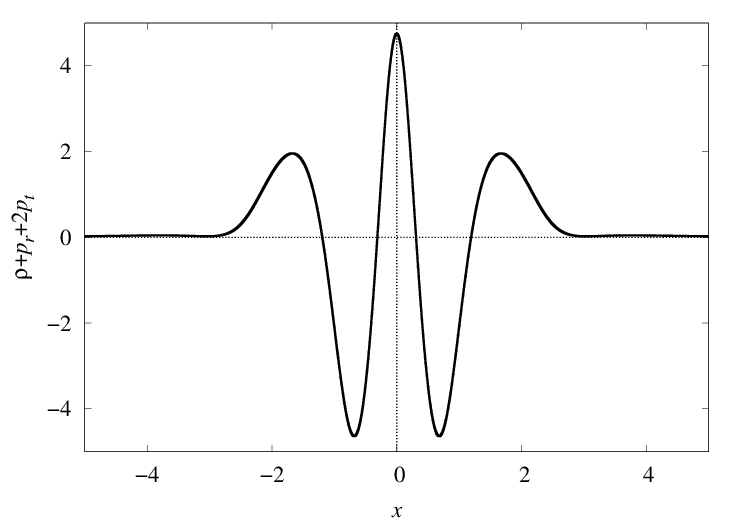}
    \caption{Graphical representation of the energy conditions for the BB model with multiple horizons and throats. The chosen parameters are $a=M=1$ and $a_0=4$.}
    \label{fig:EC}
\end{figure}

In Fig. \ref{fig:EC}, we observe how the fluid quantities combine in order to determine whether the energy conditions are violated or satisfied. It is noticeable that all combinations are symmetric under the transformation $x\rightarrow -x$ and exhibit similar oscillatory behavior. Only the $NEC_2$ and $SEC_3$ conditions are satisfied at the center, while the other conditions are violated at the center. All the energy conditions show regions where they are violated and regions where they are satisfied.

Thus, as expected for this type of solution, the violation of the energy conditions is always necessary, but not in all regions of spacetime.
\section{Conclusions and discussion}\label{S:conclusions}
In this work, we studied the possibility of the existence of BB solutions that exhibit multiple horizons and multiple throats and anti-throats. Depending on the chosen parameter values, the number of horizons and throats/anti-throats changes. The presence of throats/anti-throats is determined by analyzing the maxima and minima of the solution's area.

The analysis of the spacetime regularity is done through the Kretschmann scalar associated with our spacetime. Although we have not explicitly written the analytical expression of the Kretschmann scalar, we analyzed its asymptotic behavior and found that the curvature invariant is well-behaved both at the center of the solution and at distant points. Since the Kretschmann scalar approaches zero at infinity, the solution is asymptotically flat and recovers Schwarzschild. At the center of the solution, the invariant tends to a constant. Therefore, our solution is regular throughout the entire spacetime.

We studied the quasi-local mass of Hernandez-Misner-Sharp for the model. We found that, unlike usual BB models, the mass is not always positive. Analyzing the asymptotic limits, we obtained the same results as those found in standard BB models, $\lim_{x\rightarrow 0}M_{HMS}(x)=a/2$ and $\lim_{x\rightarrow \infty}M_{HMS}(x)=M$.

For this spacetime to be a solution of Einstein equations, it is necessary to identify the sources that generate the solution. We applied the method proposed by Bronnikov and found that the solution can be modeled by a combination of a partially phantom scalar field along with NED. The scalar field is partially phantom because the sign of the kinetic term changes depending on the position. The electromagnetic part is modeled by a magnetic monopole. The behavior of the electromagnetic Lagrangian is shown graphically since this function is obtained analytically. The potential associated with the scalar field is also analyzed graphically, displaying oscillatory behavior. The method applied to find the sources was a type of reverse engineering, where we have a metric and determine which type of source can generate this solution. However, if we use these same sources, singular solutions may appear, but by imposing certain conditions, these solutions become regular.

To study the energy conditions of the system, we considered that our system is described by an anisotropic fluid, and through the quantities of this fluid, we examined the behavior of the energy conditions. We found that all energy conditions are either violated or satisfied depending on the region of spacetime being considered. This differs from usual BB solutions, which violate the null energy condition throughout the entire spacetime, unlike in our case.

The presence of horizons can alter various properties of black holes, such as their thermodynamics, quasi-normal modes, gravitational waves, quasi-periodic oscillations, particles precession, optical appearance of the accretion disk, gravitational lens, and shadow. Therefore, in future work, we intend to investigate how these properties change when we modify the parameters of our model. We also intend to investigate this type of behavior for black holes with cylindrical symmetry. This type of solution can also be generalized to rotating black holes or dynamic solutions where the mass or the regularization parameter depends on time, similar to what is presented in \cite{Simpson:2019cer}.

\section*{Acknowledgments}
M.E.R. would like to thank Conselho Nacional de Desenvolvimento Cient\'ifico e Tecnol\'ogico - CNPq, Brazil  for partial financial support. M.S. would like to thank Funda\c c\~ao Cearense de Apoio ao Desenvolvimento Cient\'ifico e Tecnol\'ogico (FUNCAP) for partial financial support.


\bibliography{ref.bib}

\begin{thebibliography}{50}%
\makeatletter
\providecommand \@ifxundefined [1]{%
 \@ifx{#1\undefined}
}%
\providecommand \@ifnum [1]{%
 \ifnum #1\expandafter \@firstoftwo
 \else \expandafter \@secondoftwo
 \fi
}%
\providecommand \@ifx [1]{%
 \ifx #1\expandafter \@firstoftwo
 \else \expandafter \@secondoftwo
 \fi
}%
\providecommand \natexlab [1]{#1}%
\providecommand \enquote  [1]{``#1''}%
\providecommand \bibnamefont  [1]{#1}%
\providecommand \bibfnamefont [1]{#1}%
\providecommand \citenamefont [1]{#1}%
\providecommand \href@noop [0]{\@secondoftwo}%
\providecommand \href [0]{\begingroup \@sanitize@url \@href}%
\providecommand \@href[1]{\@@startlink{#1}\@@href}%
\providecommand \@@href[1]{\endgroup#1\@@endlink}%
\providecommand \@sanitize@url [0]{\catcode `\\12\catcode `\$12\catcode `\&12\catcode `\#12\catcode `\^12\catcode `\_12\catcode `\%12\relax}%
\providecommand \@@startlink[1]{}%
\providecommand \@@endlink[0]{}%
\providecommand \url  [0]{\begingroup\@sanitize@url \@url }%
\providecommand \@url [1]{\endgroup\@href {#1}{\urlprefix }}%
\providecommand \urlprefix  [0]{URL }%
\providecommand \Eprint [0]{\href }%
\providecommand \doibase [0]{http://dx.doi.org/}%
\providecommand \selectlanguage [0]{\@gobble}%
\providecommand \bibinfo  [0]{\@secondoftwo}%
\providecommand \bibfield  [0]{\@secondoftwo}%
\providecommand \translation [1]{[#1]}%
\providecommand \BibitemOpen [0]{}%
\providecommand \bibitemStop [0]{}%
\providecommand \bibitemNoStop [0]{.\EOS\space}%
\providecommand \EOS [0]{\spacefactor3000\relax}%
\providecommand \BibitemShut  [1]{\csname bibitem#1\endcsname}%
\let\auto@bib@innerbib\@empty
\bibitem [{\citenamefont {Wald}(1984)}]{Wald:1984rg}%
  \BibitemOpen
  \bibfield  {author} {\bibinfo {author} {\bibfnamefont {Robert~M.}\ \bibnamefont {Wald}},\ }\href {\doibase 10.7208/chicago/9780226870373.001.0001} {\emph {\bibinfo {title} {{General Relativity}}}}\ (\bibinfo  {publisher} {Chicago Univ. Pr.},\ \bibinfo {address} {Chicago, USA},\ \bibinfo {year} {1984})\BibitemShut {NoStop}%
\bibitem [{\citenamefont {Capozziello}\ and\ \citenamefont {De~Laurentis}(2011)}]{Capozziello:2011et}%
  \BibitemOpen
  \bibfield  {author} {\bibinfo {author} {\bibfnamefont {Salvatore}\ \bibnamefont {Capozziello}}\ and\ \bibinfo {author} {\bibfnamefont {Mariafelicia}\ \bibnamefont {De~Laurentis}},\ }\bibfield  {title} {\enquote {\bibinfo {title} {{Extended Theories of Gravity}},}\ }\href {\doibase 10.1016/j.physrep.2011.09.003} {\bibfield  {journal} {\bibinfo  {journal} {Phys. Rept.}\ }\textbf {\bibinfo {volume} {509}},\ \bibinfo {pages} {167--321} (\bibinfo {year} {2011})},\ \Eprint {http://arxiv.org/abs/1108.6266} {arXiv:1108.6266 [gr-qc]} \BibitemShut {NoStop}%
\bibitem [{\citenamefont {Ansoldi}\ and\ \citenamefont {Sindoni}(2015)}]{Ansoldi:2012qv}%
  \BibitemOpen
  \bibfield  {author} {\bibinfo {author} {\bibfnamefont {Stefano}\ \bibnamefont {Ansoldi}}\ and\ \bibinfo {author} {\bibfnamefont {Lorenzo}\ \bibnamefont {Sindoni}},\ }\bibfield  {title} {\enquote {\bibinfo {title} {{Multihorizon regular black holes}},}\ }in\ \href {\doibase 10.1142/9789814623995_0121} {\emph {\bibinfo {booktitle} {{13th Marcel Grossmann Meeting on Recent Developments in Theoretical and Experimental General Relativity, Astrophysics, and Relativistic Field Theories}}}}\ (\bibinfo {year} {2015})\ pp.\ \bibinfo {pages} {1198--1200},\ \Eprint {http://arxiv.org/abs/1209.3950} {arXiv:1209.3950 [gr-qc]} \BibitemShut {NoStop}%
\bibitem [{\citenamefont {Bronnikov}\ and\ \citenamefont {Rubin}(2012)}]{Bronnikov:2012wsj}%
  \BibitemOpen
  \bibfield  {author} {\bibinfo {author} {\bibfnamefont {Kirill~A.}\ \bibnamefont {Bronnikov}}\ and\ \bibinfo {author} {\bibfnamefont {Sergey~G.}\ \bibnamefont {Rubin}},\ }\href {\doibase 10.1142/12186} {\emph {\bibinfo {title} {{Black Holes, Cosmology and Extra Dimensions}}}}\ (\bibinfo  {publisher} {WSP},\ \bibinfo {year} {2012})\BibitemShut {NoStop}%
\bibitem [{\citenamefont {Olmo}\ and\ \citenamefont {Rubiera-Garcia}(2015)}]{Olmo:2015axa}%
  \BibitemOpen
  \bibfield  {author} {\bibinfo {author} {\bibfnamefont {Gonzalo~J.}\ \bibnamefont {Olmo}}\ and\ \bibinfo {author} {\bibfnamefont {Diego}\ \bibnamefont {Rubiera-Garcia}},\ }\bibfield  {title} {\enquote {\bibinfo {title} {{Nonsingular Black Holes in $f(R)$ Theories}},}\ }\href {\doibase 10.3390/universe1020173} {\bibfield  {journal} {\bibinfo  {journal} {Universe}\ }\textbf {\bibinfo {volume} {1}},\ \bibinfo {pages} {173--185} (\bibinfo {year} {2015})},\ \Eprint {http://arxiv.org/abs/1509.02430} {arXiv:1509.02430 [hep-th]} \BibitemShut {NoStop}%
\bibitem [{\citenamefont {Bronnikov}\ and\ \citenamefont {Fabris}(2006)}]{Bronnikov:2005gm}%
  \BibitemOpen
  \bibfield  {author} {\bibinfo {author} {\bibfnamefont {K.~A.}\ \bibnamefont {Bronnikov}}\ and\ \bibinfo {author} {\bibfnamefont {J.~C.}\ \bibnamefont {Fabris}},\ }\bibfield  {title} {\enquote {\bibinfo {title} {{Regular phantom black holes}},}\ }\href {\doibase 10.1103/PhysRevLett.96.251101} {\bibfield  {journal} {\bibinfo  {journal} {Phys. Rev. Lett.}\ }\textbf {\bibinfo {volume} {96}},\ \bibinfo {pages} {251101} (\bibinfo {year} {2006})},\ \Eprint {http://arxiv.org/abs/gr-qc/0511109} {arXiv:gr-qc/0511109} \BibitemShut {NoStop}%
\bibitem [{\citenamefont {Ansoldi}(2008)}]{Ansoldi:2008jw}%
  \BibitemOpen
  \bibfield  {author} {\bibinfo {author} {\bibfnamefont {Stefano}\ \bibnamefont {Ansoldi}},\ }\bibfield  {title} {\enquote {\bibinfo {title} {{Spherical black holes with regular center: A Review of existing models including a recent realization with Gaussian sources}},}\ }in\ \href@noop {} {\emph {\bibinfo {booktitle} {{Conference on Black Holes and Naked Singularities}}}}\ (\bibinfo {year} {2008})\ \Eprint {http://arxiv.org/abs/0802.0330} {arXiv:0802.0330 [gr-qc]} \BibitemShut {NoStop}%
\bibitem [{\citenamefont {Ayon-Beato}\ and\ \citenamefont {Garcia}(2000)}]{Ayon-Beato:2000mjt}%
  \BibitemOpen
  \bibfield  {author} {\bibinfo {author} {\bibfnamefont {Eloy}\ \bibnamefont {Ayon-Beato}}\ and\ \bibinfo {author} {\bibfnamefont {Alberto}\ \bibnamefont {Garcia}},\ }\bibfield  {title} {\enquote {\bibinfo {title} {{The Bardeen model as a nonlinear magnetic monopole}},}\ }\href {\doibase 10.1016/S0370-2693(00)01125-4} {\bibfield  {journal} {\bibinfo  {journal} {Phys. Lett. B}\ }\textbf {\bibinfo {volume} {493}},\ \bibinfo {pages} {149--152} (\bibinfo {year} {2000})},\ \Eprint {http://arxiv.org/abs/gr-qc/0009077} {arXiv:gr-qc/0009077} \BibitemShut {NoStop}%
\bibitem [{\citenamefont {Simpson}\ and\ \citenamefont {Visser}(2019{\natexlab{a}})}]{Simpson:2019mud}%
  \BibitemOpen
  \bibfield  {author} {\bibinfo {author} {\bibfnamefont {Alex}\ \bibnamefont {Simpson}}\ and\ \bibinfo {author} {\bibfnamefont {Matt}\ \bibnamefont {Visser}},\ }\bibfield  {title} {\enquote {\bibinfo {title} {{Regular black holes with asymptotically Minkowski cores}},}\ }\href {\doibase 10.3390/universe6010008} {\bibfield  {journal} {\bibinfo  {journal} {Universe}\ }\textbf {\bibinfo {volume} {6}},\ \bibinfo {pages} {8} (\bibinfo {year} {2019}{\natexlab{a}})},\ \Eprint {http://arxiv.org/abs/1911.01020} {arXiv:1911.01020 [gr-qc]} \BibitemShut {NoStop}%
\bibitem [{\citenamefont {Bronnikov}(2024)}]{Bronnikov:2024izh}%
  \BibitemOpen
  \bibfield  {author} {\bibinfo {author} {\bibfnamefont {Kirill~A.}\ \bibnamefont {Bronnikov}},\ }\bibfield  {title} {\enquote {\bibinfo {title} {{Regular black holes as an alternative to black bounce}},}\ }\href {\doibase 10.1103/PhysRevD.110.024021} {\bibfield  {journal} {\bibinfo  {journal} {Phys. Rev. D}\ }\textbf {\bibinfo {volume} {110}},\ \bibinfo {pages} {024021} (\bibinfo {year} {2024})},\ \Eprint {http://arxiv.org/abs/2404.14816} {arXiv:2404.14816 [gr-qc]} \BibitemShut {NoStop}%
\bibitem [{\citenamefont {Bolokhov}\ \emph {et~al.}(2024)\citenamefont {Bolokhov}, \citenamefont {Bronnikov},\ and\ \citenamefont {Skvortsova}}]{Bolokhov:2024sdy}%
  \BibitemOpen
  \bibfield  {author} {\bibinfo {author} {\bibfnamefont {S.~V.}\ \bibnamefont {Bolokhov}}, \bibinfo {author} {\bibfnamefont {K.~A.}\ \bibnamefont {Bronnikov}}, \ and\ \bibinfo {author} {\bibfnamefont {M.~V.}\ \bibnamefont {Skvortsova}},\ }\bibfield  {title} {\enquote {\bibinfo {title} {{A Regular Center Instead of a Black Bounce}},}\ }\href {\doibase 10.1134/S0202289324700178} {\bibfield  {journal} {\bibinfo  {journal} {Grav. Cosmol.}\ }\textbf {\bibinfo {volume} {30}},\ \bibinfo {pages} {265--278} (\bibinfo {year} {2024})},\ \Eprint {http://arxiv.org/abs/2405.09124} {arXiv:2405.09124 [gr-qc]} \BibitemShut {NoStop}%
\bibitem [{\citenamefont {Simpson}\ and\ \citenamefont {Visser}(2019{\natexlab{b}})}]{Simpson:2018tsi}%
  \BibitemOpen
  \bibfield  {author} {\bibinfo {author} {\bibfnamefont {Alex}\ \bibnamefont {Simpson}}\ and\ \bibinfo {author} {\bibfnamefont {Matt}\ \bibnamefont {Visser}},\ }\bibfield  {title} {\enquote {\bibinfo {title} {{Black-bounce to traversable wormhole}},}\ }\href {\doibase 10.1088/1475-7516/2019/02/042} {\bibfield  {journal} {\bibinfo  {journal} {JCAP}\ }\textbf {\bibinfo {volume} {02}},\ \bibinfo {pages} {042} (\bibinfo {year} {2019}{\natexlab{b}})},\ \Eprint {http://arxiv.org/abs/1812.07114} {arXiv:1812.07114 [gr-qc]} \BibitemShut {NoStop}%
\bibitem [{\citenamefont {Lobo}\ \emph {et~al.}(2021)\citenamefont {Lobo}, \citenamefont {Rodrigues}, \citenamefont {de~Sousa~Silva}, \citenamefont {Simpson},\ and\ \citenamefont {Visser}}]{Lobo:2020ffi}%
  \BibitemOpen
  \bibfield  {author} {\bibinfo {author} {\bibfnamefont {Francisco S.~N.}\ \bibnamefont {Lobo}}, \bibinfo {author} {\bibfnamefont {Manuel~E.}\ \bibnamefont {Rodrigues}}, \bibinfo {author} {\bibfnamefont {Marcos~V.}\ \bibnamefont {de~Sousa~Silva}}, \bibinfo {author} {\bibfnamefont {Alex}\ \bibnamefont {Simpson}}, \ and\ \bibinfo {author} {\bibfnamefont {Matt}\ \bibnamefont {Visser}},\ }\bibfield  {title} {\enquote {\bibinfo {title} {{Novel black-bounce spacetimes: wormholes, regularity, energy conditions, and causal structure}},}\ }\href {\doibase 10.1103/PhysRevD.103.084052} {\bibfield  {journal} {\bibinfo  {journal} {Phys. Rev. D}\ }\textbf {\bibinfo {volume} {103}},\ \bibinfo {pages} {084052} (\bibinfo {year} {2021})},\ \Eprint {http://arxiv.org/abs/2009.12057} {arXiv:2009.12057 [gr-qc]} \BibitemShut {NoStop}%
\bibitem [{\citenamefont {Franzin}\ \emph {et~al.}(2021)\citenamefont {Franzin}, \citenamefont {Liberati}, \citenamefont {Mazza}, \citenamefont {Simpson},\ and\ \citenamefont {Visser}}]{Franzin:2021vnj}%
  \BibitemOpen
  \bibfield  {author} {\bibinfo {author} {\bibfnamefont {Edgardo}\ \bibnamefont {Franzin}}, \bibinfo {author} {\bibfnamefont {Stefano}\ \bibnamefont {Liberati}}, \bibinfo {author} {\bibfnamefont {Jacopo}\ \bibnamefont {Mazza}}, \bibinfo {author} {\bibfnamefont {Alex}\ \bibnamefont {Simpson}}, \ and\ \bibinfo {author} {\bibfnamefont {Matt}\ \bibnamefont {Visser}},\ }\bibfield  {title} {\enquote {\bibinfo {title} {{Charged black-bounce spacetimes}},}\ }\href {\doibase 10.1088/1475-7516/2021/07/036} {\bibfield  {journal} {\bibinfo  {journal} {JCAP}\ }\textbf {\bibinfo {volume} {07}},\ \bibinfo {pages} {036} (\bibinfo {year} {2021})},\ \Eprint {http://arxiv.org/abs/2104.11376} {arXiv:2104.11376 [gr-qc]} \BibitemShut {NoStop}%
\bibitem [{\citenamefont {Furtado}\ and\ \citenamefont {Alencar}(2022)}]{Furtado:2022tnb}%
  \BibitemOpen
  \bibfield  {author} {\bibinfo {author} {\bibfnamefont {Job}\ \bibnamefont {Furtado}}\ and\ \bibinfo {author} {\bibfnamefont {Geov\'a}\ \bibnamefont {Alencar}},\ }\bibfield  {title} {\enquote {\bibinfo {title} {{BTZ Black-Bounce to Traversable Wormhole}},}\ }\href {\doibase 10.3390/universe8120625} {\bibfield  {journal} {\bibinfo  {journal} {Universe}\ }\textbf {\bibinfo {volume} {8}},\ \bibinfo {pages} {625} (\bibinfo {year} {2022})},\ \Eprint {http://arxiv.org/abs/2210.06608} {arXiv:2210.06608 [gr-qc]} \BibitemShut {NoStop}%
\bibitem [{\citenamefont {Lima}\ \emph {et~al.}(2023{\natexlab{a}})\citenamefont {Lima}, \citenamefont {de~Alencar~Filho},\ and\ \citenamefont {Furtado~Neto}}]{Lima:2022pvc}%
  \BibitemOpen
  \bibfield  {author} {\bibinfo {author} {\bibfnamefont {Arthur~Menezes}\ \bibnamefont {Lima}}, \bibinfo {author} {\bibfnamefont {Geov\'a~Maciel}\ \bibnamefont {de~Alencar~Filho}}, \ and\ \bibinfo {author} {\bibfnamefont {Job~Saraiva}\ \bibnamefont {Furtado~Neto}},\ }\bibfield  {title} {\enquote {\bibinfo {title} {{Black String Bounce to Traversable Wormhole}},}\ }\href {\doibase 10.3390/sym15010150} {\bibfield  {journal} {\bibinfo  {journal} {Symmetry}\ }\textbf {\bibinfo {volume} {15}},\ \bibinfo {pages} {150} (\bibinfo {year} {2023}{\natexlab{a}})},\ \Eprint {http://arxiv.org/abs/2211.12349} {arXiv:2211.12349 [gr-qc]} \BibitemShut {NoStop}%
\bibitem [{\citenamefont {Bronnikov}\ \emph {et~al.}(2023)\citenamefont {Bronnikov}, \citenamefont {Rodrigues},\ and\ \citenamefont {de~S.~Silva}}]{Bronnikov:2023aya}%
  \BibitemOpen
  \bibfield  {author} {\bibinfo {author} {\bibfnamefont {Kirill~A.}\ \bibnamefont {Bronnikov}}, \bibinfo {author} {\bibfnamefont {Manuel~E.}\ \bibnamefont {Rodrigues}}, \ and\ \bibinfo {author} {\bibfnamefont {Marcos~V.}\ \bibnamefont {de~S.~Silva}},\ }\bibfield  {title} {\enquote {\bibinfo {title} {{Cylindrical black bounces and their field sources}},}\ }\href {\doibase 10.1103/PhysRevD.108.024065} {\bibfield  {journal} {\bibinfo  {journal} {Phys. Rev. D}\ }\textbf {\bibinfo {volume} {108}},\ \bibinfo {pages} {024065} (\bibinfo {year} {2023})},\ \Eprint {http://arxiv.org/abs/2305.19296} {arXiv:2305.19296 [gr-qc]} \BibitemShut {NoStop}%
\bibitem [{\citenamefont {Lima}\ \emph {et~al.}(2023{\natexlab{b}})\citenamefont {Lima}, \citenamefont {Alencar}, \citenamefont {Costa~Filho},\ and\ \citenamefont {Landim}}]{Lima:2023arg}%
  \BibitemOpen
  \bibfield  {author} {\bibinfo {author} {\bibfnamefont {A.}~\bibnamefont {Lima}}, \bibinfo {author} {\bibfnamefont {G.}~\bibnamefont {Alencar}}, \bibinfo {author} {\bibfnamefont {R.~N.}\ \bibnamefont {Costa~Filho}}, \ and\ \bibinfo {author} {\bibfnamefont {R.~R.}\ \bibnamefont {Landim}},\ }\bibfield  {title} {\enquote {\bibinfo {title} {{Charged black string bounce and its field source}},}\ }\href {\doibase 10.1007/s10714-023-03156-x} {\bibfield  {journal} {\bibinfo  {journal} {Gen. Rel. Grav.}\ }\textbf {\bibinfo {volume} {55}},\ \bibinfo {pages} {108} (\bibinfo {year} {2023}{\natexlab{b}})},\ \Eprint {http://arxiv.org/abs/2306.03029} {arXiv:2306.03029 [gr-qc]} \BibitemShut {NoStop}%
\bibitem [{\citenamefont {Lima}\ \emph {et~al.}(2024)\citenamefont {Lima}, \citenamefont {Alencar},\ and\ \citenamefont {S\'aez-Chillon~G\'omez}}]{Lima:2023jtl}%
  \BibitemOpen
  \bibfield  {author} {\bibinfo {author} {\bibfnamefont {A.}~\bibnamefont {Lima}}, \bibinfo {author} {\bibfnamefont {G.}~\bibnamefont {Alencar}}, \ and\ \bibinfo {author} {\bibfnamefont {Diego}\ \bibnamefont {S\'aez-Chillon~G\'omez}},\ }\bibfield  {title} {\enquote {\bibinfo {title} {{Regularizing rotating black strings with a new black-bounce solution}},}\ }\href {\doibase 10.1103/PhysRevD.109.064038} {\bibfield  {journal} {\bibinfo  {journal} {Phys. Rev. D}\ }\textbf {\bibinfo {volume} {109}},\ \bibinfo {pages} {064038} (\bibinfo {year} {2024})},\ \Eprint {http://arxiv.org/abs/2307.07404} {arXiv:2307.07404 [gr-qc]} \BibitemShut {NoStop}%
\bibitem [{\citenamefont {Crispim}\ \emph {et~al.}(2024)\citenamefont {Crispim}, \citenamefont {Estrada}, \citenamefont {Muniz},\ and\ \citenamefont {Alencar}}]{Crispim:2024yjz}%
  \BibitemOpen
  \bibfield  {author} {\bibinfo {author} {\bibfnamefont {Tiago~M.}\ \bibnamefont {Crispim}}, \bibinfo {author} {\bibfnamefont {Milko}\ \bibnamefont {Estrada}}, \bibinfo {author} {\bibfnamefont {C.~R.}\ \bibnamefont {Muniz}}, \ and\ \bibinfo {author} {\bibfnamefont {G.}~\bibnamefont {Alencar}},\ }\bibfield  {title} {\enquote {\bibinfo {title} {{Braneworld black bounce to transversable wormhole}},}\ }\href {\doibase 10.1088/1475-7516/2024/10/063} {\bibfield  {journal} {\bibinfo  {journal} {JCAP}\ }\textbf {\bibinfo {volume} {10}},\ \bibinfo {pages} {063} (\bibinfo {year} {2024})},\ \Eprint {http://arxiv.org/abs/2405.08048} {arXiv:2405.08048 [hep-th]} \BibitemShut {NoStop}%
\bibitem [{\citenamefont {Lima}\ \emph {et~al.}(2021)\citenamefont {Lima}, \citenamefont {Crispino}, \citenamefont {Cunha},\ and\ \citenamefont {Herdeiro}}]{Lima:2021las}%
  \BibitemOpen
  \bibfield  {author} {\bibinfo {author} {\bibfnamefont {Haroldo C.~D.}\ \bibnamefont {Lima}, \bibfnamefont {Junior.}}, \bibinfo {author} {\bibfnamefont {Lu\'\i{}s C.~B.}\ \bibnamefont {Crispino}}, \bibinfo {author} {\bibfnamefont {Pedro V.~P.}\ \bibnamefont {Cunha}}, \ and\ \bibinfo {author} {\bibfnamefont {Carlos A.~R.}\ \bibnamefont {Herdeiro}},\ }\bibfield  {title} {\enquote {\bibinfo {title} {{Can different black holes cast the same shadow?}}}\ }\href {\doibase 10.1103/PhysRevD.103.084040} {\bibfield  {journal} {\bibinfo  {journal} {Phys. Rev. D}\ }\textbf {\bibinfo {volume} {103}},\ \bibinfo {pages} {084040} (\bibinfo {year} {2021})},\ \Eprint {http://arxiv.org/abs/2102.07034} {arXiv:2102.07034 [gr-qc]} \BibitemShut {NoStop}%
\bibitem [{\citenamefont {Guerrero}\ \emph {et~al.}(2021)\citenamefont {Guerrero}, \citenamefont {Olmo}, \citenamefont {Rubiera-Garcia},\ and\ \citenamefont {G\'omez}}]{Guerrero:2021ues}%
  \BibitemOpen
  \bibfield  {author} {\bibinfo {author} {\bibfnamefont {Merce}\ \bibnamefont {Guerrero}}, \bibinfo {author} {\bibfnamefont {Gonzalo~J.}\ \bibnamefont {Olmo}}, \bibinfo {author} {\bibfnamefont {Diego}\ \bibnamefont {Rubiera-Garcia}}, \ and\ \bibinfo {author} {\bibfnamefont {Diego S\'aez-Chill\'on}\ \bibnamefont {G\'omez}},\ }\bibfield  {title} {\enquote {\bibinfo {title} {{Shadows and optical appearance of black bounces illuminated by a thin accretion disk}},}\ }\href {\doibase 10.1088/1475-7516/2021/08/036} {\bibfield  {journal} {\bibinfo  {journal} {JCAP}\ }\textbf {\bibinfo {volume} {08}},\ \bibinfo {pages} {036} (\bibinfo {year} {2021})},\ \Eprint {http://arxiv.org/abs/2105.15073} {arXiv:2105.15073 [gr-qc]} \BibitemShut {NoStop}%
\bibitem [{\citenamefont {Vagnozzi}\ \emph {et~al.}(2023)\citenamefont {Vagnozzi} \emph {et~al.}}]{Vagnozzi:2022moj}%
  \BibitemOpen
  \bibfield  {author} {\bibinfo {author} {\bibfnamefont {Sunny}\ \bibnamefont {Vagnozzi}} \emph {et~al.},\ }\bibfield  {title} {\enquote {\bibinfo {title} {{Horizon-scale tests of gravity theories and fundamental physics from the Event Horizon Telescope image of Sagittarius A}},}\ }\href {\doibase 10.1088/1361-6382/acd97b} {\bibfield  {journal} {\bibinfo  {journal} {Class. Quant. Grav.}\ }\textbf {\bibinfo {volume} {40}},\ \bibinfo {pages} {165007} (\bibinfo {year} {2023})},\ \Eprint {http://arxiv.org/abs/2205.07787} {arXiv:2205.07787 [gr-qc]} \BibitemShut {NoStop}%
\bibitem [{\citenamefont {Bronnikov}\ and\ \citenamefont {Walia}(2022)}]{Bronnikov:2021uta}%
  \BibitemOpen
  \bibfield  {author} {\bibinfo {author} {\bibfnamefont {Kirill~A.}\ \bibnamefont {Bronnikov}}\ and\ \bibinfo {author} {\bibfnamefont {Rahul~Kumar}\ \bibnamefont {Walia}},\ }\bibfield  {title} {\enquote {\bibinfo {title} {{Field sources for Simpson-Visser spacetimes}},}\ }\href {\doibase 10.1103/PhysRevD.105.044039} {\bibfield  {journal} {\bibinfo  {journal} {Phys. Rev. D}\ }\textbf {\bibinfo {volume} {105}},\ \bibinfo {pages} {044039} (\bibinfo {year} {2022})},\ \Eprint {http://arxiv.org/abs/2112.13198} {arXiv:2112.13198 [gr-qc]} \BibitemShut {NoStop}%
\bibitem [{\citenamefont {Bronnikov}(2022{\natexlab{a}})}]{Bronnikov:2022bud}%
  \BibitemOpen
  \bibfield  {author} {\bibinfo {author} {\bibfnamefont {K.~A.}\ \bibnamefont {Bronnikov}},\ }\bibfield  {title} {\enquote {\bibinfo {title} {{Black bounces, wormholes, and partly phantom scalar fields}},}\ }\href {\doibase 10.1103/PhysRevD.106.064029} {\bibfield  {journal} {\bibinfo  {journal} {Phys. Rev. D}\ }\textbf {\bibinfo {volume} {106}},\ \bibinfo {pages} {064029} (\bibinfo {year} {2022}{\natexlab{a}})},\ \Eprint {http://arxiv.org/abs/2206.09227} {arXiv:2206.09227 [gr-qc]} \BibitemShut {NoStop}%
\bibitem [{\citenamefont {Ma}\ and\ \citenamefont {Zhao}(2014)}]{Ma:2014qma}%
  \BibitemOpen
  \bibfield  {author} {\bibinfo {author} {\bibfnamefont {Meng-Sen}\ \bibnamefont {Ma}}\ and\ \bibinfo {author} {\bibfnamefont {Ren}\ \bibnamefont {Zhao}},\ }\bibfield  {title} {\enquote {\bibinfo {title} {{Corrected form of the first law of thermodynamics for regular black holes}},}\ }\href {\doibase 10.1088/0264-9381/31/24/245014} {\bibfield  {journal} {\bibinfo  {journal} {Class. Quant. Grav.}\ }\textbf {\bibinfo {volume} {31}},\ \bibinfo {pages} {245014} (\bibinfo {year} {2014})},\ \Eprint {http://arxiv.org/abs/1411.0833} {arXiv:1411.0833 [gr-qc]} \BibitemShut {NoStop}%
\bibitem [{\citenamefont {Zhang}\ and\ \citenamefont {Gao}(2018)}]{Zhang:2016ilt}%
  \BibitemOpen
  \bibfield  {author} {\bibinfo {author} {\bibfnamefont {Yuan}\ \bibnamefont {Zhang}}\ and\ \bibinfo {author} {\bibfnamefont {Sijie}\ \bibnamefont {Gao}},\ }\bibfield  {title} {\enquote {\bibinfo {title} {{First law and Smarr formula of black hole mechanics in nonlinear gauge theories}},}\ }\href {\doibase 10.1088/1361-6382/aac9d4} {\bibfield  {journal} {\bibinfo  {journal} {Class. Quant. Grav.}\ }\textbf {\bibinfo {volume} {35}},\ \bibinfo {pages} {145007} (\bibinfo {year} {2018})},\ \Eprint {http://arxiv.org/abs/1610.01237} {arXiv:1610.01237 [gr-qc]} \BibitemShut {NoStop}%
\bibitem [{\citenamefont {Maluf}\ and\ \citenamefont {Neves}(2018)}]{Maluf:2018lyu}%
  \BibitemOpen
  \bibfield  {author} {\bibinfo {author} {\bibfnamefont {R.~V.}\ \bibnamefont {Maluf}}\ and\ \bibinfo {author} {\bibfnamefont {Juliano C.~S.}\ \bibnamefont {Neves}},\ }\bibfield  {title} {\enquote {\bibinfo {title} {{Thermodynamics of a class of regular black holes with a generalized uncertainty principle}},}\ }\href {\doibase 10.1103/PhysRevD.97.104015} {\bibfield  {journal} {\bibinfo  {journal} {Phys. Rev. D}\ }\textbf {\bibinfo {volume} {97}},\ \bibinfo {pages} {104015} (\bibinfo {year} {2018})},\ \Eprint {http://arxiv.org/abs/1801.02661} {arXiv:1801.02661 [gr-qc]} \BibitemShut {NoStop}%
\bibitem [{\citenamefont {Rodrigues}\ \emph {et~al.}(2022)\citenamefont {Rodrigues}, \citenamefont {de~S.~Silva},\ and\ \citenamefont {Vieira}}]{Rodrigues:2022qdp}%
  \BibitemOpen
  \bibfield  {author} {\bibinfo {author} {\bibfnamefont {Manuel~E.}\ \bibnamefont {Rodrigues}}, \bibinfo {author} {\bibfnamefont {Marcos~V.}\ \bibnamefont {de~S.~Silva}}, \ and\ \bibinfo {author} {\bibfnamefont {Henrique~A.}\ \bibnamefont {Vieira}},\ }\bibfield  {title} {\enquote {\bibinfo {title} {{Bardeen-Kiselev black hole with a cosmological constant}},}\ }\href {\doibase 10.1103/PhysRevD.105.084043} {\bibfield  {journal} {\bibinfo  {journal} {Phys. Rev. D}\ }\textbf {\bibinfo {volume} {105}},\ \bibinfo {pages} {084043} (\bibinfo {year} {2022})},\ \Eprint {http://arxiv.org/abs/2203.04965} {arXiv:2203.04965 [gr-qc]} \BibitemShut {NoStop}%
\bibitem [{\citenamefont {de~S.~Silva}(2024)}]{deSSilva:2024fmp}%
  \BibitemOpen
  \bibfield  {author} {\bibinfo {author} {\bibfnamefont {Marcos~V.}\ \bibnamefont {de~S.~Silva}},\ }\bibfield  {title} {\enquote {\bibinfo {title} {{Charged Black Hole with Inverse Electrodynamics}},}\ }\href {\doibase 10.1007/s10773-024-05760-2} {\bibfield  {journal} {\bibinfo  {journal} {Int. J. Theor. Phys.}\ }\textbf {\bibinfo {volume} {63}},\ \bibinfo {pages} {220} (\bibinfo {year} {2024})}\BibitemShut {NoStop}%
\bibitem [{\citenamefont {Bronnikov}(2001)}]{Bronnikov:2000vy}%
  \BibitemOpen
  \bibfield  {author} {\bibinfo {author} {\bibfnamefont {Kirill~A.}\ \bibnamefont {Bronnikov}},\ }\bibfield  {title} {\enquote {\bibinfo {title} {{Regular magnetic black holes and monopoles from nonlinear electrodynamics}},}\ }\href {\doibase 10.1103/PhysRevD.63.044005} {\bibfield  {journal} {\bibinfo  {journal} {Phys. Rev. D}\ }\textbf {\bibinfo {volume} {63}},\ \bibinfo {pages} {044005} (\bibinfo {year} {2001})},\ \Eprint {http://arxiv.org/abs/gr-qc/0006014} {arXiv:gr-qc/0006014} \BibitemShut {NoStop}%
\bibitem [{\citenamefont {Novello}\ \emph {et~al.}(2000)\citenamefont {Novello}, \citenamefont {De~Lorenci}, \citenamefont {Salim},\ and\ \citenamefont {Klippert}}]{Novello:1999pg}%
  \BibitemOpen
  \bibfield  {author} {\bibinfo {author} {\bibfnamefont {M.}~\bibnamefont {Novello}}, \bibinfo {author} {\bibfnamefont {V.~A.}\ \bibnamefont {De~Lorenci}}, \bibinfo {author} {\bibfnamefont {J.~M.}\ \bibnamefont {Salim}}, \ and\ \bibinfo {author} {\bibfnamefont {Renato}\ \bibnamefont {Klippert}},\ }\bibfield  {title} {\enquote {\bibinfo {title} {{Geometrical aspects of light propagation in nonlinear electrodynamics}},}\ }\href {\doibase 10.1103/PhysRevD.61.045001} {\bibfield  {journal} {\bibinfo  {journal} {Phys. Rev. D}\ }\textbf {\bibinfo {volume} {61}},\ \bibinfo {pages} {045001} (\bibinfo {year} {2000})},\ \Eprint {http://arxiv.org/abs/gr-qc/9911085} {arXiv:gr-qc/9911085} \BibitemShut {NoStop}%
\bibitem [{\citenamefont {Toshmatov}\ \emph {et~al.}(2021)\citenamefont {Toshmatov}, \citenamefont {Ahmedov},\ and\ \citenamefont {Malafarina}}]{Toshmatov:2021fgm}%
  \BibitemOpen
  \bibfield  {author} {\bibinfo {author} {\bibfnamefont {Bobir}\ \bibnamefont {Toshmatov}}, \bibinfo {author} {\bibfnamefont {Bobomurat}\ \bibnamefont {Ahmedov}}, \ and\ \bibinfo {author} {\bibfnamefont {Daniele}\ \bibnamefont {Malafarina}},\ }\bibfield  {title} {\enquote {\bibinfo {title} {{Can a light ray distinguish charge of a black hole in nonlinear electrodynamics?}}}\ }\href {\doibase 10.1103/PhysRevD.103.024026} {\bibfield  {journal} {\bibinfo  {journal} {Phys. Rev. D}\ }\textbf {\bibinfo {volume} {103}},\ \bibinfo {pages} {024026} (\bibinfo {year} {2021})},\ \Eprint {http://arxiv.org/abs/2101.05496} {arXiv:2101.05496 [gr-qc]} \BibitemShut {NoStop}%
\bibitem [{\citenamefont {Bronnikov}(2022{\natexlab{b}})}]{Bronnikov:2022ofk}%
  \BibitemOpen
  \bibfield  {author} {\bibinfo {author} {\bibfnamefont {Kirill~A.}\ \bibnamefont {Bronnikov}},\ }\bibfield  {title} {\enquote {\bibinfo {title} {{Regular black holes sourced by nonlinear electrodynamics}},}\ }\href@noop {} {\  (\bibinfo {year} {2022}{\natexlab{b}})},\ \Eprint {http://arxiv.org/abs/2211.00743} {arXiv:2211.00743 [gr-qc]} \BibitemShut {NoStop}%
\bibitem [{\citenamefont {Silva}\ and\ \citenamefont {Rodrigues}(2024)}]{Silva:2024fpn}%
  \BibitemOpen
  \bibfield  {author} {\bibinfo {author} {\bibfnamefont {Marcos V. de~S.}\ \bibnamefont {Silva}}\ and\ \bibinfo {author} {\bibfnamefont {Manuel~E.}\ \bibnamefont {Rodrigues}},\ }\bibfield  {title} {\enquote {\bibinfo {title} {{Orbits Around a Black Bounce Spacetime}},}\ }\href {\doibase 10.1007/s10773-024-05644-5} {\bibfield  {journal} {\bibinfo  {journal} {Int. J. Theor. Phys.}\ }\textbf {\bibinfo {volume} {63}},\ \bibinfo {pages} {101} (\bibinfo {year} {2024})},\ \Eprint {http://arxiv.org/abs/2404.15792} {arXiv:2404.15792 [gr-qc]} \BibitemShut {NoStop}%
\bibitem [{\citenamefont {Stuchl\'\i{}k}\ and\ \citenamefont {Schee}(2019)}]{Stuchlik:2019uvf}%
  \BibitemOpen
  \bibfield  {author} {\bibinfo {author} {\bibfnamefont {Zdenek}\ \bibnamefont {Stuchl\'\i{}k}}\ and\ \bibinfo {author} {\bibfnamefont {Jan}\ \bibnamefont {Schee}},\ }\bibfield  {title} {\enquote {\bibinfo {title} {{Shadow of the regular Bardeen black holes and comparison of the motion of photons and neutrinos}},}\ }\href {\doibase 10.1140/epjc/s10052-019-6543-8} {\bibfield  {journal} {\bibinfo  {journal} {Eur. Phys. J. C}\ }\textbf {\bibinfo {volume} {79}},\ \bibinfo {pages} {44} (\bibinfo {year} {2019})}\BibitemShut {NoStop}%
\bibitem [{\citenamefont {Rayimbaev}\ \emph {et~al.}(2022)\citenamefont {Rayimbaev}, \citenamefont {Bardiev}, \citenamefont {Mirzaev}, \citenamefont {Abdujabbarov},\ and\ \citenamefont {Khalmirzaev}}]{Rayimbaev:2022znx}%
  \BibitemOpen
  \bibfield  {author} {\bibinfo {author} {\bibfnamefont {Javlon}\ \bibnamefont {Rayimbaev}}, \bibinfo {author} {\bibfnamefont {Dilshodbek}\ \bibnamefont {Bardiev}}, \bibinfo {author} {\bibfnamefont {Temurbek}\ \bibnamefont {Mirzaev}}, \bibinfo {author} {\bibfnamefont {Ahmadjon}\ \bibnamefont {Abdujabbarov}}, \ and\ \bibinfo {author} {\bibfnamefont {Akram}\ \bibnamefont {Khalmirzaev}},\ }\bibfield  {title} {\enquote {\bibinfo {title} {{Shadow and massless particles around regular Bardeen black holes in 4D Einstein Gauss\textendash{}Bonnet gravity}},}\ }\href {\doibase 10.1142/S0218271822500559} {\bibfield  {journal} {\bibinfo  {journal} {Int. J. Mod. Phys. D}\ }\textbf {\bibinfo {volume} {31}},\ \bibinfo {pages} {2250055} (\bibinfo {year} {2022})}\BibitemShut {NoStop}%
\bibitem [{\citenamefont {de~Paula}\ \emph {et~al.}(2023)\citenamefont {de~Paula}, \citenamefont {Lima~Junior}, \citenamefont {Cunha},\ and\ \citenamefont {Crispino}}]{dePaula:2023ozi}%
  \BibitemOpen
  \bibfield  {author} {\bibinfo {author} {\bibfnamefont {Marco A.~A.}\ \bibnamefont {de~Paula}}, \bibinfo {author} {\bibfnamefont {Haroldo C.~D.}\ \bibnamefont {Lima~Junior}}, \bibinfo {author} {\bibfnamefont {Pedro V.~P.}\ \bibnamefont {Cunha}}, \ and\ \bibinfo {author} {\bibfnamefont {Lu\'\i{}s C.~B.}\ \bibnamefont {Crispino}},\ }\bibfield  {title} {\enquote {\bibinfo {title} {{Electrically charged regular black holes in nonlinear electrodynamics: Light rings, shadows, and gravitational lensing}},}\ }\href {\doibase 10.1103/PhysRevD.108.084029} {\bibfield  {journal} {\bibinfo  {journal} {Phys. Rev. D}\ }\textbf {\bibinfo {volume} {108}},\ \bibinfo {pages} {084029} (\bibinfo {year} {2023})},\ \Eprint {http://arxiv.org/abs/2305.04776} {arXiv:2305.04776 [gr-qc]} \BibitemShut {NoStop}%
\bibitem [{\citenamefont {da~Silva}\ \emph {et~al.}(2023)\citenamefont {da~Silva}, \citenamefont {Lobo}, \citenamefont {Olmo},\ and\ \citenamefont {Rubiera-Garcia}}]{daSilva:2023jxa}%
  \BibitemOpen
  \bibfield  {author} {\bibinfo {author} {\bibfnamefont {Lu\'\i{}s F.~Dias}\ \bibnamefont {da~Silva}}, \bibinfo {author} {\bibfnamefont {Francisco S.~N.}\ \bibnamefont {Lobo}}, \bibinfo {author} {\bibfnamefont {Gonzalo~J.}\ \bibnamefont {Olmo}}, \ and\ \bibinfo {author} {\bibfnamefont {Diego}\ \bibnamefont {Rubiera-Garcia}},\ }\bibfield  {title} {\enquote {\bibinfo {title} {{Photon rings as tests for alternative spherically symmetric geometries with thin accretion disks}},}\ }\href {\doibase 10.1103/PhysRevD.108.084055} {\bibfield  {journal} {\bibinfo  {journal} {Phys. Rev. D}\ }\textbf {\bibinfo {volume} {108}},\ \bibinfo {pages} {084055} (\bibinfo {year} {2023})},\ \Eprint {http://arxiv.org/abs/2307.06778} {arXiv:2307.06778 [gr-qc]} \BibitemShut {NoStop}%
\bibitem [{\citenamefont {Nojiri}\ and\ \citenamefont {Odintsov}(2017)}]{Nojiri:2017kex}%
  \BibitemOpen
  \bibfield  {author} {\bibinfo {author} {\bibfnamefont {Shin'ichi}\ \bibnamefont {Nojiri}}\ and\ \bibinfo {author} {\bibfnamefont {S.~D.}\ \bibnamefont {Odintsov}},\ }\bibfield  {title} {\enquote {\bibinfo {title} {{Regular multihorizon black holes in modified gravity with nonlinear electrodynamics}},}\ }\href {\doibase 10.1103/PhysRevD.96.104008} {\bibfield  {journal} {\bibinfo  {journal} {Phys. Rev. D}\ }\textbf {\bibinfo {volume} {96}},\ \bibinfo {pages} {104008} (\bibinfo {year} {2017})},\ \Eprint {http://arxiv.org/abs/1708.05226} {arXiv:1708.05226 [hep-th]} \BibitemShut {NoStop}%
\bibitem [{\citenamefont {Gao}\ \emph {et~al.}(2018)\citenamefont {Gao}, \citenamefont {Lu}, \citenamefont {Yu},\ and\ \citenamefont {Shen}}]{Gao:2017vqv}%
  \BibitemOpen
  \bibfield  {author} {\bibinfo {author} {\bibfnamefont {Changjun}\ \bibnamefont {Gao}}, \bibinfo {author} {\bibfnamefont {Youjun}\ \bibnamefont {Lu}}, \bibinfo {author} {\bibfnamefont {Shuang}\ \bibnamefont {Yu}}, \ and\ \bibinfo {author} {\bibfnamefont {You-Gen}\ \bibnamefont {Shen}},\ }\bibfield  {title} {\enquote {\bibinfo {title} {{Black hole and cosmos with multiple horizons and multiple singularities in vector-tensor theories}},}\ }\href {\doibase 10.1103/PhysRevD.97.104013} {\bibfield  {journal} {\bibinfo  {journal} {Phys. Rev. D}\ }\textbf {\bibinfo {volume} {97}},\ \bibinfo {pages} {104013} (\bibinfo {year} {2018})},\ \Eprint {http://arxiv.org/abs/1711.00996} {arXiv:1711.00996 [gr-qc]} \BibitemShut {NoStop}%
\bibitem [{\citenamefont {Rodrigues}\ and\ \citenamefont {de~Sousa~Silva}(2019)}]{Rodrigues:2019xrc}%
  \BibitemOpen
  \bibfield  {author} {\bibinfo {author} {\bibfnamefont {Manuel~E.}\ \bibnamefont {Rodrigues}}\ and\ \bibinfo {author} {\bibfnamefont {Marcos~V.}\ \bibnamefont {de~Sousa~Silva}},\ }\bibfield  {title} {\enquote {\bibinfo {title} {{Regular multihorizon black holes in $f(G)$ gravity with nonlinear electrodynamics}},}\ }\href {\doibase 10.1103/PhysRevD.99.124010} {\bibfield  {journal} {\bibinfo  {journal} {Phys. Rev. D}\ }\textbf {\bibinfo {volume} {99}},\ \bibinfo {pages} {124010} (\bibinfo {year} {2019})},\ \Eprint {http://arxiv.org/abs/1906.06168} {arXiv:1906.06168 [gr-qc]} \BibitemShut {NoStop}%
\bibitem [{\citenamefont {Rodrigues}\ \emph {et~al.}(2020)\citenamefont {Rodrigues}, \citenamefont {de~Sousa~Silva},\ and\ \citenamefont {de~Siqueira}}]{Rodrigues:2020pem}%
  \BibitemOpen
  \bibfield  {author} {\bibinfo {author} {\bibfnamefont {Manuel~E.}\ \bibnamefont {Rodrigues}}, \bibinfo {author} {\bibfnamefont {Marcos~V.}\ \bibnamefont {de~Sousa~Silva}}, \ and\ \bibinfo {author} {\bibfnamefont {Andrew~S.}\ \bibnamefont {de~Siqueira}},\ }\bibfield  {title} {\enquote {\bibinfo {title} {{Regular multihorizon black holes in General Relativity}},}\ }\href {\doibase 10.1103/PhysRevD.102.084038} {\bibfield  {journal} {\bibinfo  {journal} {Phys. Rev. D}\ }\textbf {\bibinfo {volume} {102}},\ \bibinfo {pages} {084038} (\bibinfo {year} {2020})},\ \Eprint {http://arxiv.org/abs/2010.09490} {arXiv:2010.09490 [gr-qc]} \BibitemShut {NoStop}%
\bibitem [{\citenamefont {Nashed}\ and\ \citenamefont {Nojiri}(2021)}]{Nashed:2021ctg}%
  \BibitemOpen
  \bibfield  {author} {\bibinfo {author} {\bibfnamefont {G.~G.~L.}\ \bibnamefont {Nashed}}\ and\ \bibinfo {author} {\bibfnamefont {Shin'ichi}\ \bibnamefont {Nojiri}},\ }\bibfield  {title} {\enquote {\bibinfo {title} {{Mimetic Euler-Heisenberg theory, charged solutions, and multihorizon black holes}},}\ }\href {\doibase 10.1103/PhysRevD.104.044043} {\bibfield  {journal} {\bibinfo  {journal} {Phys. Rev. D}\ }\textbf {\bibinfo {volume} {104}},\ \bibinfo {pages} {044043} (\bibinfo {year} {2021})},\ \Eprint {http://arxiv.org/abs/2107.13550} {arXiv:2107.13550 [gr-qc]} \BibitemShut {NoStop}%
\bibitem [{\citenamefont {Morris}\ and\ \citenamefont {Thorne}(1988)}]{Morris:1988cz}%
  \BibitemOpen
  \bibfield  {author} {\bibinfo {author} {\bibfnamefont {M.~S.}\ \bibnamefont {Morris}}\ and\ \bibinfo {author} {\bibfnamefont {K.~S.}\ \bibnamefont {Thorne}},\ }\bibfield  {title} {\enquote {\bibinfo {title} {{Wormholes in space-time and their use for interstellar travel: A tool for teaching general relativity}},}\ }\href {\doibase 10.1119/1.15620} {\bibfield  {journal} {\bibinfo  {journal} {Am. J. Phys.}\ }\textbf {\bibinfo {volume} {56}},\ \bibinfo {pages} {395--412} (\bibinfo {year} {1988})}\BibitemShut {NoStop}%
\bibitem [{\citenamefont {Vishveshwara}(1968)}]{Vishveshwara1968}%
  \BibitemOpen
  \bibfield  {author} {\bibinfo {author} {\bibfnamefont {C.~V.}\ \bibnamefont {Vishveshwara}},\ }\bibfield  {title} {\enquote {\bibinfo {title} {{Generalization of the ``Schwarzschild Surface''' to Arbitrary Static and Stationary Metrics}},}\ }\href {\doibase 10.1063/1.1664717} {\bibfield  {journal} {\bibinfo  {journal} {J. Math. Phys.}\ }\textbf {\bibinfo {volume} {9}},\ \bibinfo {pages} {1319--1322} (\bibinfo {year} {1968})}\BibitemShut {NoStop}%
\bibitem [{\citenamefont {Rodrigues}\ and\ \citenamefont {Silva}(2023)}]{Rodrigues:2022mdm}%
  \BibitemOpen
  \bibfield  {author} {\bibinfo {author} {\bibfnamefont {Manuel~E.}\ \bibnamefont {Rodrigues}}\ and\ \bibinfo {author} {\bibfnamefont {Marcos V. de~S.}\ \bibnamefont {Silva}},\ }\bibfield  {title} {\enquote {\bibinfo {title} {{Black-bounces with multiple throats and anti-throats}},}\ }\href {\doibase 10.1088/1361-6382/ad0195} {\bibfield  {journal} {\bibinfo  {journal} {Class. Quant. Grav.}\ }\textbf {\bibinfo {volume} {40}},\ \bibinfo {pages} {225011} (\bibinfo {year} {2023})},\ \Eprint {http://arxiv.org/abs/2204.11851} {arXiv:2204.11851 [gr-qc]} \BibitemShut {NoStop}%
\bibitem [{\citenamefont {Visser}(1995)}]{Visser:1995cc}%
  \BibitemOpen
  \bibfield  {author} {\bibinfo {author} {\bibfnamefont {Matt}\ \bibnamefont {Visser}},\ }\href@noop {} {\emph {\bibinfo {title} {{Lorentzian wormholes: From Einstein to Hawking}}}}\ (\bibinfo {year} {1995})\BibitemShut {NoStop}%
\bibitem [{\citenamefont {Alencar}\ \emph {et~al.}(2024)\citenamefont {Alencar}, \citenamefont {Bronnikov}, \citenamefont {Rodrigues}, \citenamefont {S\'aez-Chill\'on~G\'omez},\ and\ \citenamefont {de~S.~Silva}}]{Alencar:2024yvh}%
  \BibitemOpen
  \bibfield  {author} {\bibinfo {author} {\bibfnamefont {G.}~\bibnamefont {Alencar}}, \bibinfo {author} {\bibfnamefont {Kirill~A.}\ \bibnamefont {Bronnikov}}, \bibinfo {author} {\bibfnamefont {Manuel~E.}\ \bibnamefont {Rodrigues}}, \bibinfo {author} {\bibfnamefont {Diego}\ \bibnamefont {S\'aez-Chill\'on~G\'omez}}, \ and\ \bibinfo {author} {\bibfnamefont {Marcos~V.}\ \bibnamefont {de~S.~Silva}},\ }\bibfield  {title} {\enquote {\bibinfo {title} {{On black bounce space-times in non-linear electrodynamics}},}\ }\href {\doibase 10.1140/epjc/s10052-024-13119-4} {\bibfield  {journal} {\bibinfo  {journal} {Eur. Phys. J. C}\ }\textbf {\bibinfo {volume} {84}},\ \bibinfo {pages} {745} (\bibinfo {year} {2024})},\ \Eprint {http://arxiv.org/abs/2403.12897} {arXiv:2403.12897 [gr-qc]} \BibitemShut {NoStop}%
\bibitem [{\citenamefont {Simpson}\ \emph {et~al.}(2019)\citenamefont {Simpson}, \citenamefont {Martin-Moruno},\ and\ \citenamefont {Visser}}]{Simpson:2019cer}%
  \BibitemOpen
  \bibfield  {author} {\bibinfo {author} {\bibfnamefont {Alex}\ \bibnamefont {Simpson}}, \bibinfo {author} {\bibfnamefont {Prado}\ \bibnamefont {Martin-Moruno}}, \ and\ \bibinfo {author} {\bibfnamefont {Matt}\ \bibnamefont {Visser}},\ }\bibfield  {title} {\enquote {\bibinfo {title} {{Vaidya spacetimes, black-bounces, and traversable wormholes}},}\ }\href {\doibase 10.1088/1361-6382/ab28a5} {\bibfield  {journal} {\bibinfo  {journal} {Class. Quant. Grav.}\ }\textbf {\bibinfo {volume} {36}},\ \bibinfo {pages} {145007} (\bibinfo {year} {2019})},\ \Eprint {http://arxiv.org/abs/1902.04232} {arXiv:1902.04232 [gr-qc]} \BibitemShut {NoStop}%
\end{thebibliography}%
\end{document}